\documentclass[aps,prd,amsmath,floats,floatfix,superscriptaddress,nofootinbib,showpacs,onecolumn]{revtex4}

\usepackage{amssymb}
\usepackage{amsmath}
\usepackage{verbatim}
\usepackage{mathrsfs}
\usepackage{amsfonts}
\usepackage{latexsym}
\usepackage{epsfig}
\usepackage{color}
\usepackage{graphicx}
\usepackage{units}
\usepackage{overpic}
\usepackage[hidelinks]{hyperref}
\usepackage{afterpage}
\usepackage{float}
\usepackage{mathtools}
\usepackage{xcolor}
\hypersetup{
    colorlinks,
    linkcolor={red!50!black},
    citecolor={blue!50!black},
    urlcolor={blue!80!black}
}

\usepackage{booktabs, multirow} 
\usepackage{soul}
\usepackage{changepage,threeparttable} 
\begin{document}


\definecolor{orange}{rgb}{0.9,0.45,0} 
\definecolor{applegreen}{rgb}{0.55, 0.71, 0.0}
\definecolor{blue}{rgb}{0.0,0.0,1.0} 

\newcommand{\vjp}[1]{{\textcolor{orange}{[VJ: #1]}}}
\newcommand{\dmc}[1]{{\textcolor{blue}{[DM: #1]}}}
\newcommand{\juanc}[1]{{\textcolor{green}{[JCD: #1]}}}
\newcommand{\dario}[1]{{\textcolor{red}{[Dario: #1]}}}


\title{Born-Infeld boson stars}

\author{V\'ictor Jaramillo}
\affiliation{Instituto de Ciencias Nucleares, Universidad Nacional Aut\'onoma de M\'exico,
Circuito Exterior C.U., A.P. 70-543, M\'exico D.F. 04510, M\'exico}

\author{Daniel Mart\'inez-Carbajal}
\affiliation{Instituto de Ciencias Nucleares, Universidad Nacional Aut\'onoma de M\'exico,
Circuito Exterior C.U., A.P. 70-543, M\'exico D.F. 04510, M\'exico}

\author{Juan Carlos Degollado} 
\affiliation{Instituto de Ciencias F\'isicas, Universidad Nacional Aut\'onoma 
de M\'exico, Apartado Postal 48-3, 62251, Cuernavaca, Morelos, M\'exico}

\author{Dar\'io N\'u\~nez}
\affiliation{Instituto de Ciencias Nucleares, Universidad Nacional Aut\'onoma de M\'exico,
Circuito Exterior C.U., A.P. 70-543, M\'exico D.F. 04510, M\'exico}


\date{\today}

\begin{abstract}
We study the Einstein-Klein-Gordon system coupled to the Born-Infeld electrodynamics.  
We explore the solution space of a static spherically symmetric, complex scalar field 
minimally coupled to both gravitational and  electromagnetic fields. The resulting 
asymptotically flat solutions resemble the known charged boson stars in Maxwell electrodynamics.
The behaviour of such configurations as a function of the Born-Infeld parameter $b$
and the scalar field charge parameter $q$ has been analyzed.
Given $b$, a critical value for $q$ exists beyond which no static solutions exist, 
we find that the value of this critical charge increases with respect to the Maxwell case ($b\to\infty$) as $b$ decreases.
We also found that Born-Infeld boson stars have lower mass for any finite value of the Born-Infeld parameter and that their compactness is lower than Maxwell's counterparts.

\end{abstract}




\maketitle


\section{Introduction}
\label{Sec:intro}

The Born-Infeld (BI) theory of electrodynamics was proposed in order to remove the infinity in the potential energy of the electron that occurs in the Coulomb field of the Maxwell theory \cite{Born:1934gh}. Born and Infeld imposed a maximum of the strength of the electromagnetic field at the origin of a point charge, in analogy to the action principle of free particles in relativistic theories where naturally impose an upper bound on the velocity of particles \cite{Chernitsky:1999ra,Jacksonbook:14242,Kruglov:2016uzf}.
This non-linear generalization of Maxwell’s electrodynamics was followed by Heisenberg and Euler who presented an effective action describing non-linear corrections to Maxwell’s theory due to one-loop effects in the electron-positron interaction \cite{2006physics...5038H}. 
Ever since, other models of non-linear electrodynamics have been extensively studied in many areas of theoretical physics including strong gravity \cite{BeltranJimenez:2017doy,Odintsov:2014yaa}, string theory \cite{Maldacena:1997re} and cosmology \cite{Jana:2016uvq,BeltranJimenez:2017uwv}.
A common scenario for gravity related applications of nonlinear electrodynamics is to use a particular model as a matter source in Einstein's field equations.  
In this sense, several models of nonlinear electrodynamics have been used to produce regular solutions that possess event horizons \cite{Ayon-Beato:1998hmi, Ayon-Beato:1999kuh,Ayon-Beato:2000mjt,Bronnikov:2000vy,Konoplya:2011qq,Toshmatov:2015wga,Kruglov:2017mpj} and, in particular Einstein BI gravity has been considered to describe charged black holes \cite{Babar:2021nst,Jafarzade:2020ova, Ali:2022zox, Falciano:2021kdu, Gan:2019jac}. 

The fact that different electrodynamical theories lead to different charged black hole solutions, raises the question
of whether other compact solutions in General Relativity exist. In this work we use the BI theory coupled to
a complex scalar field to produce self gravitating regular objects. Similar solutions, known as charged boson stars have
been found previously by coupling Maxwell’s electrodynamics to a complex boson field see for instance~\cite{Jetzer:1989av,Kleihaus:2009kr,Pugliese:2013gsa,Kan:2017rqk,Collodel:2019ohy} and more recently \cite{Lopez:2023phk}.
In this work, we have found stationary  BI boson stars by solving numerically the Einstein-Born-Infeld-scalar field system in spherical symmetry.

Charged boson stars share many properties of their non charged counterparts, both families are characterized by the frequency of the field and there is a critical mass that separates between stable and unstable configurations \cite{Kaup:1968zz, Ruffini.187.1767, Jetzer:1989us, Jetzer:1991jr,Schunck:2003kk,Liebling:2012fv}. 
The charged case however, includes an extra parameter, namely the charge of the particle, that allows for a larger family of solutions. Furthermore, there is a critical charge above which solutions ceases to exist, this is related with the fact that electromagnetic interaction may be larger than the gravitational one. For larger values of the charge the gravitational binding energy is not enough to balance the electromagnetic repulsion and the configuration tears apart.
In this work we analyze BI boson stars  as a function of the BI parameter $b$ and determine some of their properties. We found that the solutions in both electromagnetic theories are qualitatively equivalent with some important quantitative differences in global quantities such as the total mass, charge and compactness. The differences are more evident in the regime where $b$ is of the order of unity or less. This result is consistent with the fact that the BI field reduces to Maxwell's in the limit of $b\rightarrow \infty$.
The work is organized as follows: In Section \ref{Sec:model} we introduce the model and derive the field equations, we also provide the boundary conditions for the fields in order to get asymptotically flat regular solutions and some useful diagnostic quantities. In Section \ref{Sec:BIfield} we present some solutions and describe their properties. Finally in Section \ref{Sec:Conclusions} we give some conclusions. Throughout this work we use units with $c=G=1$
and the mostly plus metric signature $(-,+,+,+)$.

\section{Charged boson stars}
\label{Sec:model}

\subsection{Field equations}
We consider the theory of a complex scalar field $\Phi$, minimally coupled to Einstein’s gravity and with the BI electrodynamics. The action is given by 
\begin{equation}\label{eq: EBIaction}
S=\int d^4 x\sqrt{-g} \left[\frac{1}{16\pi }\mathcal{R}-\frac{1}{2}\left(g^{\mu\nu}(D_\mu\Phi)(D_\nu\Phi)^*+\mu^2|\Phi|^2+\frac{\lambda}{2}|\Phi|^{4}\right) + {\cal L}_{\rm BI}(F)
\right],
\end{equation}
where $\mathcal{R}$ is the Ricci scalar, $\mu$ is the scalar field particle mass, $\lambda$ is the coupling constant, $A_\mu$ is the gauge field, $q$ is the boson charge and $D_\mu=\nabla_\mu+iqA_\mu$ is the covariant derivative. The BI Lagrangian ${\cal L}_{\rm BI}(F)$, is given by
\begin{equation}\label{eq:}
{\cal L}_{\rm BI}(F)=b^{2} \left(1-\sqrt{1+\frac{F}{2b^{2}}}\right) ,
\end{equation}
where $F=F_{\mu\nu}F^{\mu\nu}$, $F_{\mu\nu}=\partial_\mu A_\nu- 
 \partial_\nu A_\mu$ and $b$ is the BI parameter. 
In the limit $b \rightarrow \infty$ the  Maxwell’s electrodynamics ${\cal L}_{\rm Max}(F)= -F/4$, is recovered. The Einstein–BI-scalar equations can be obtained by varying the action Eq.~(\ref{eq: EBIaction}) with respect to the different fields  leading to corresponding field equations (see \textit{e.g.} \cite{Hawking:1973uf}).
The variation of the action with respect to the metric tensor $g_{\mu\nu}$ leads to 
\begin{subequations}\label{eq:ekgm}
\begin{eqnarray}
&& R_{\mu\nu}-\frac{1}{2}\mathcal{R} g_{\mu\nu}=8\pi T_{\mu\nu} \ ,
\label{eq:einstein} \\
&& T_{\mu\nu}=T^{\Phi}_{\mu\nu}+{T^{\mathrm{BI}}}_{\mu\nu} \ ,
\end{eqnarray}
\end{subequations}
where the stress energy tensors are
\begin{eqnarray}
\label{Eq:tmunu}
&& T^{\Phi}_{\mu\nu}=\frac{1}{2}(D
_{\mu}\Phi)(D_{\nu}\Phi)^*+\frac{1}{2}(D_{\nu}\Phi)(D_{\mu}\Phi)^*
-\frac{1}{2}g_{\mu\nu}\left(g^{\alpha\beta}(D_{\alpha}\Phi)(D_{\beta}\Phi)^*+\mu^2|\Phi|^2+\frac{\lambda}{2}|\Phi|^{4}\right),\\
&& {T^{\mathrm{BI}}}_{\mu\nu}=\left[\frac{F_{\mu\sigma} F_{\nu\lambda}g^{\sigma\lambda}}{\sqrt{1+\frac{F}{2b^{2}}}}+g_{\mu\nu}\,b^{2}\left(1-\sqrt{1+\frac{F
}{2b^{2}}}\right)\right].
\end{eqnarray}
Variation with respect to the field $\Phi$ leads to the Klein-Gordon equation,
\begin{equation}\label{eq:kg}
    g^{\mu\nu}D_\nu D_\mu \Phi=\left(\mu^2+\lambda|\Phi|^{2}\right)\Phi \ .
\end{equation}
Finally, variation with respect to the gauge field $A_{\mu}$, leads to the generalized Maxwell equations with the scalar field as source 
\begin{subequations}\label{eq:maxwell}
  \begin{eqnarray}
    \nabla_\nu \left(\frac{F^{\mu\nu}}{\sqrt{1+\frac{F}{2b^{2}}}}\right)=q j^\mu,  
  \end{eqnarray}
 where the current is given by
  \begin{eqnarray}
       j^\mu=\frac{ig^{\mu\nu}}{2}[\Phi^*D_\nu\Phi-\Phi(D_\nu{\Phi_{}})^*] \ .
       \label{eq:current}
  \end{eqnarray}
\end{subequations}
%

\subsection{Spherically symetric spacetime and Ans\"atze for the fields}

We are interested in static spherically symmetric solutions satisfying the following ansatz in isotropic coordinates
\begin{equation}\label{eq:metric}
  g_{\mu\nu}dx^\mu dx^\nu=-N^2\, dt^2 + \Psi^4\left(dr^2+r^2 d\,\Omega^2\right),
\end{equation}
where $N, \Psi$ are functions of $r$ and $d\Omega^{2}$ is the metric defined on the unitary two sphere.
Besides spherical symmetry,
we furthermore assume that the scalar field $\Phi$, has a harmonic time dependence of the form   
\begin{equation}\label{eq:ansatzphi}
\Phi=\phi(r)e^{-i\omega t }\ ,
\end{equation}
where $\omega$ is a real constant.
In spherical symmetry, the gauge potential $A_\mu$ is written as %
\begin{equation}\label{eq:ansatzA}
  A_\mu dx^\mu=C(r) dt \ . 
\end{equation}
This choice guarantees that any magnetic field vanishes. 

Using the ansatz for the metric \eqref{eq:metric} and for the fields~(\ref{eq:ansatzphi}-\ref{eq:ansatzA})
into the equations.~(\ref{eq:ekgm}-\ref{eq:maxwell}) yield the following: \\
Einstein equations \footnote{The equations displayed correspond to the mixed $tt$ component and to the trace of Einstein's equations respectively. }
\begin{eqnarray}
    &&\Delta_3\Psi+\pi\Psi^5\left[\left(\frac{(qC-\omega)\phi}{N}\right)^2+\frac{\partial\phi^2}{\Psi^4}+    \left(\mu^2+\frac{\lambda}{2}\phi^2\right) \phi^2-2{b}^2 \left(1 - \frac{1}{\sqrt{1-\left(\frac{\partial C}{\Psi^2Nb}\right)^2}}\right)\right]=0\ ,\label{IsoEinstein}\\
    &&\Delta_3 N+\frac{2\partial N\partial\Psi}{\Psi}-4\pi N\Psi^4\left[2\left(\frac{(qC-\omega)\phi}{N}\right)^2-    \left(\mu^2+\frac{\lambda}{2}\phi^2\right) \phi^2+2b^2\left(1-\sqrt{1-\left(\frac{\partial C}{\Psi^2Nb}\right)^2}\right)\right]=0 \ ,\label{IsoBI}
\end{eqnarray}
scalar field equations
\begin{equation}\label{eq:kg_eq}
    \Delta_3\phi+\frac{\partial \phi\partial N}{N}+2\frac{\partial \phi\partial \Psi}{\Psi}-\Psi^4\left(\mu^2+ \lambda\phi^2-\left(\frac{qC-\omega}{N}\right)^2\right)\phi=0\ ,
\end{equation}
and Born-Infeld field equations
\begin{equation}
\Delta_3 C+\frac{2\partial C\partial\Psi}{\Psi}-\frac{\partial C\partial N}{N}
    -2\frac{{\partial C}^3}{N^2{b}^2\Psi^5}\left(\frac{\Psi}{r}+2\partial\Psi \right)
    -q\Psi^4\left( 1-\left(\frac{\partial C}{\Psi^2N b}\right)^2\right)^{3/2}(qC-\omega)\phi^2=0 \ ,
    \label{eq:CEBI}
\end{equation}
where $\Delta_{3}:=\frac{d^{2}}{dr^{2}}+\frac{2}{r}\frac{d}{dr}$ and
we have used the shorthand 
notation $\partial f:=\frac{df}{dr}$. 
The equations for neutral boson stars are obtained in the limit $q\rightarrow 0$.

In order to solve the system of differential equations for the functions $\{\phi, C,N,\Psi\}$ and the unknown parameter $\omega$, one has to impose boundary conditions on the scalar field, the gauge potential and the metric functions. 
Static solutions are uniquely specified by the central value of the boson field $\phi(0)$ and the behaviour at infinity.
Imposing regularity of the functions at the origin leads to
\begin{equation}\label{eq:regularity_r}
  \begin{split}
    &\phi|_{r=0}=0, \quad \partial C|_{r= 0}=0;\\
    &\partial N |_{r=0}=0, \quad \partial \Psi |_{r=0}=0 \ .
  \end{split}
\end{equation}
whereas asymptotic flatness implies 
\begin{equation}\label{eq:out_bc}
  \begin{split}
    &N|_{r\to\infty}=1, \quad \Psi|_{r\to\infty}=1; \\
    &\phi|_{r\to\infty}=0,\quad C|_{r\to\infty}=0 \ . 
  \end{split}
\end{equation}
Additionally, the condition $\omega<\mu$ is necessary in order to guarantee asymptotically flat solutions.
%
\subsection{Diagnostics quantities}

For an stationary and asymptotically flat spacetime, the total mass can be computed using the Komar expression 
\cite{Wald:1984rg}
\begin{equation}\label{eq:KomarM}
  M_{\rm K}=\frac{1}{4\pi}\int_{\Sigma_t}R_{\mu\nu}n^\mu\xi^\nu dV \ ,
\end{equation}
where $\Sigma_t$  denotes a spacelike hypersurface, $n^{\mu}$ is a timelike vector normal to $\Sigma_t$ and $\xi=\partial_{t}$ is the timelike Killing vector. 
The Komar expression \eqref{eq:KomarM} with the metric \eqref{eq:metric}, reduces in the limit $r\rightarrow \infty$ to $M_{\rm K}=\lim_{r\to\infty}r^2\partial N$. 

The total mass of the system can also be determined from the ADM (Arnowitt-Desser-Misner) mass definition as given in Eq. (4.20) of reference \cite{Gourgoulhon:2010ju}, 
which, given the stationarity and asymptotic flatness of the spacetime must coincide with the Komar mass \cite{1979JMP....20..793A,Gourgoulhon:2010ju}. For the metric (\ref{eq:metric}), the ADM mass can be computed from the expression 
\begin{equation}
M_{\rm ADM}=-2\lim_{r\to\infty}r^2 \partial \Psi \ .
\end{equation}
We compute the Komar and ADM masses and when they coincide up to a relative difference of less than $10^{-5}$, we consider the numerical code has converged to the desired solution with a mass $M_{\rm K} = M_{\rm ADM} = M$. \\
The action Eq.~(\ref{eq: EBIaction}) 
is invariant under the global $U(1)$ transformations $\Phi\rightarrow\Phi e^{i\alpha},$
this implies that the current~(\ref{eq:current}) is conserved and satisfies the conservation law $\nabla_{\mu}j^{\mu}=0$. Integration of the conserved law over a spacelike hypersurface $\Sigma_t$ defines the conserved Noether charge
\begin{equation}\label{KomarN}
  \mathcal{N}=\int_{\Sigma_t} j^\mu n_\mu dV \ ,
\end{equation}
which can be associated with the total number of bosonic particles \cite{Ruffini.187.1767}. Furthermore, one can define the charge of the configuration as $Q=q \mathcal{N}$. After replacing the current Eq.~(\ref{eq:current}) in Eq.~(\ref{KomarN}), we obtain 
\begin{equation}
\mathcal{N}=4\pi\intop_{0}^{\infty}\Psi^{6}r^{2}\frac{\phi_{0}^{2}}{N}\left(Cq-\omega\right)\,dr \ . \label{Npart}
\end{equation}
Following Ref.~\cite{Jetzer:1989av}, we define the radius of the boson stars as 
\begin{equation}
R_s=\frac{1}{\mathcal{N}}\int_{\Sigma_{t}}(\Psi^2 r)\,j^{t}n_{t}\,dV=\frac{4\pi}{\mathcal{N}}\intop_{0}^{\infty}\Psi^{8}r^{3}\frac{\phi_{0}^{2}}{N}\left(Cq-\omega\right)\,dr \ .\label{Rad}
\end{equation}

Additionally, the energy density of the scalar field measured by a static observer is calculated as $\rho=-T^0{}_{0}$. This quantity thus has contributions of the scalar field and of the BI field
\begin{equation}\label{eq:density}
    \rho = \rho_{\Phi}+\rho_{BI}
\end{equation}
where
\begin{equation}
\rho_{\Phi}=\frac{1}{2}\left[\frac{(qC-\omega)^2\phi^2}{N^2}+\frac{\partial\phi^2}{\Psi^4}+    \left(\mu^2+\frac{\lambda}{2}\phi^2\right) \phi^2\right]\,, \qquad 
\rho_{BI}=-{b}^2 \left(1 - \frac{1}{\sqrt{1-\left(\frac{\partial C}{\Psi^2Nb}\right)^2}}\right) \ .
\end{equation}
Finally, we define the compactness of the stars as the ratio
\begin{equation}
    \mathcal{C}= \frac{M}{R_s} \ ,
\end{equation}
where the effective size of the boson stars
$R_s$, is defined in Eq.~\eqref{Rad}.

\section{Results}
\label{Sec:BIfield}
%
The non-linear system (\ref{IsoEinstein}-\ref{eq:CEBI}) is solved numerically with boundary conditions (\ref{eq:out_bc}) and (\ref{eq:regularity_r}) using a spectral collocation method with Chebyshev polynomials as spectral basis in a compactified domain, see for instance Ref.~\cite{Grandclement:2007sb}. The solutions are found by means of a Newton-Raphson iteration. In this work only ground state solutions (with no nodes for $\phi(r>0)$) are considered. We took neutral boson stars solutions, constructed in \cite{Alcubierre:2021psa}, as starting point to generate charged solutions with fixed $b\gg1$ and slowly increasing the value of $q$, following the same procedure as in \cite{Jaramillo:2022gcq}. Then, we looked for solutions with smaller values of $b$ and fixed $q$. Finally, for each pair $(q,b)$ the sequence of solutions was constructed carefully varying the value of the lapse function $N$ at $r=0$ or equivalently 
$\phi_0=\phi(r=0)$.

We use the invariance of the equations
(\ref{IsoEinstein}-\ref{eq:CEBI})
under the scaling 
\begin{equation}\label{eq:tilde}
    r \rightarrow  \tilde r = \mu r , \quad \omega \rightarrow  \tilde \omega = \omega/\mu, \quad 
    \lambda   \rightarrow  \tilde \lambda = \lambda/\mu^2 , \quad
    b   \rightarrow  \tilde b = b/\mu^2, 
\end{equation}
to obtain solutions for arbitrary values of $\mu$ from those presented below. 
Furthermore, in order to facilitate a comparison with the charge and scalar field scale reported in the literature, \textit{e.g.} \cite{Jetzer:1989av,Kleihaus:2009kr,Kain:2021bwd} we use the following variables
\begin{equation}
    \tilde{q}=\frac{q}{\mu \sqrt{8\pi}},\quad\tilde{\phi}=\sqrt{4 \pi}\phi \ .
\end{equation}

As a first approach, we consider the cases with $\lambda=0$.
A non relativistic analysis in Maxwell electrodynamics suggests that solutions only exist for
$\tilde{q}<q_{\rm crit}=1/\sqrt{2}\approx 0.707$, since otherwise the Coulomb repulsion 
could exceed the gravitational attraction \cite{Jetzer:1989av}.
This value for the maximum charge 
not necessarily holds in the relativistic case nor in BI electrodynamics. 
For instance, in the BI case,
for the sequence of solutions with fixed values
$(\tilde{b},\tilde{\phi}_0)=(0.01, 0.5)$, equilibrium solutions exist up to a maximal value 
$\tilde{q} \approx 0.86$,
while for $(\tilde{b},\tilde{\phi}_0)=(0.001, 0.5)$,
the maximum value allowed, after which the code
fails to converge, is $\tilde{q}\approx 1.01$. 
Exploring different sequences of solutions, we deduce that the maximum value of charge increases when either $\tilde{\phi}_0$
increases or $b$ decreases. 
This is related to the fact that BI electrodynamics is weaker than Maxwell's and therefore a higher value of $q$ is required to disrupt the star.
Nevertheless, in all the results presented in this section, we consider only solutions with
$\tilde{q}\leq0.7$. 
\begin{figure}[ht]
\includegraphics[width=.45\linewidth]{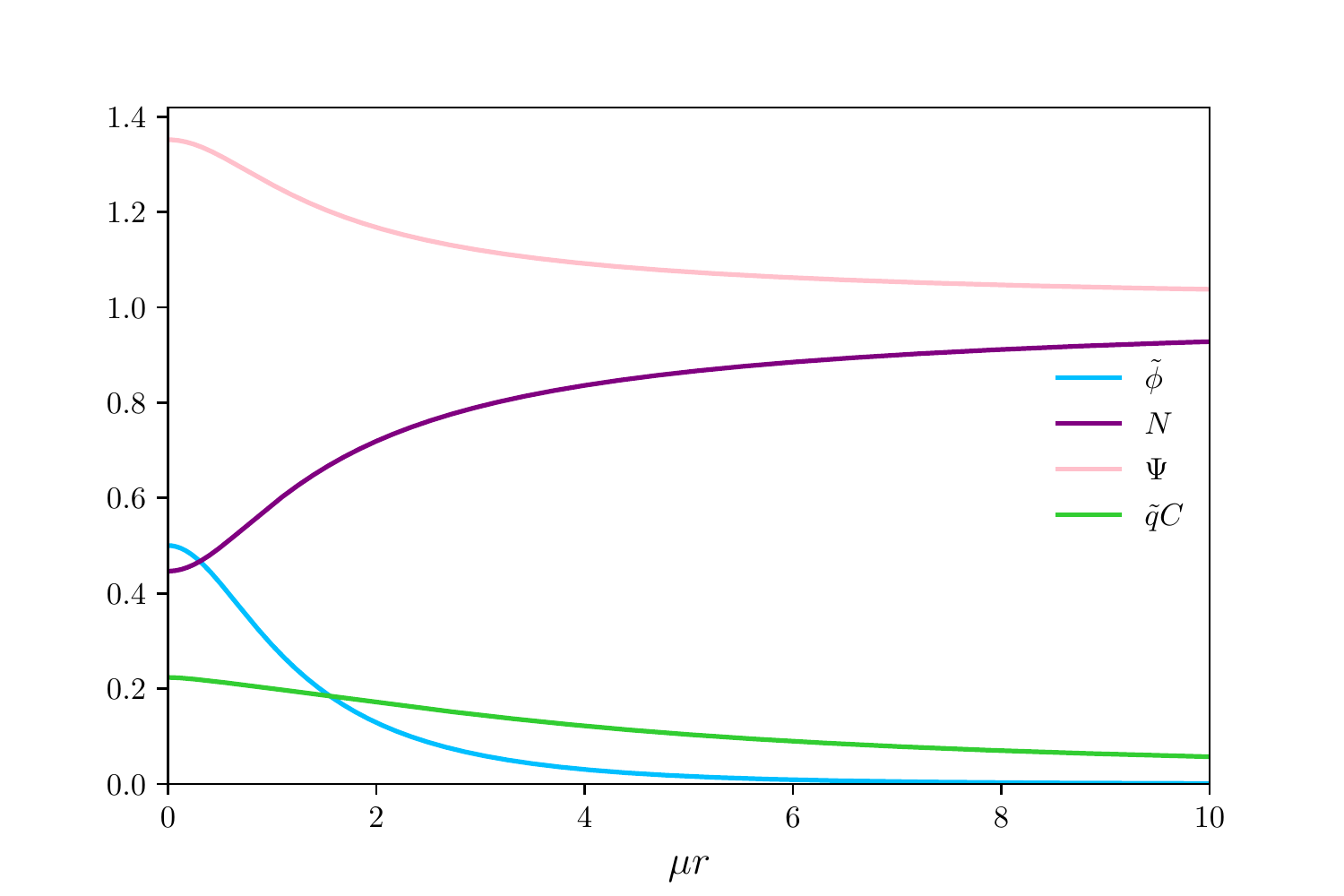}
\includegraphics[width=.45\linewidth]{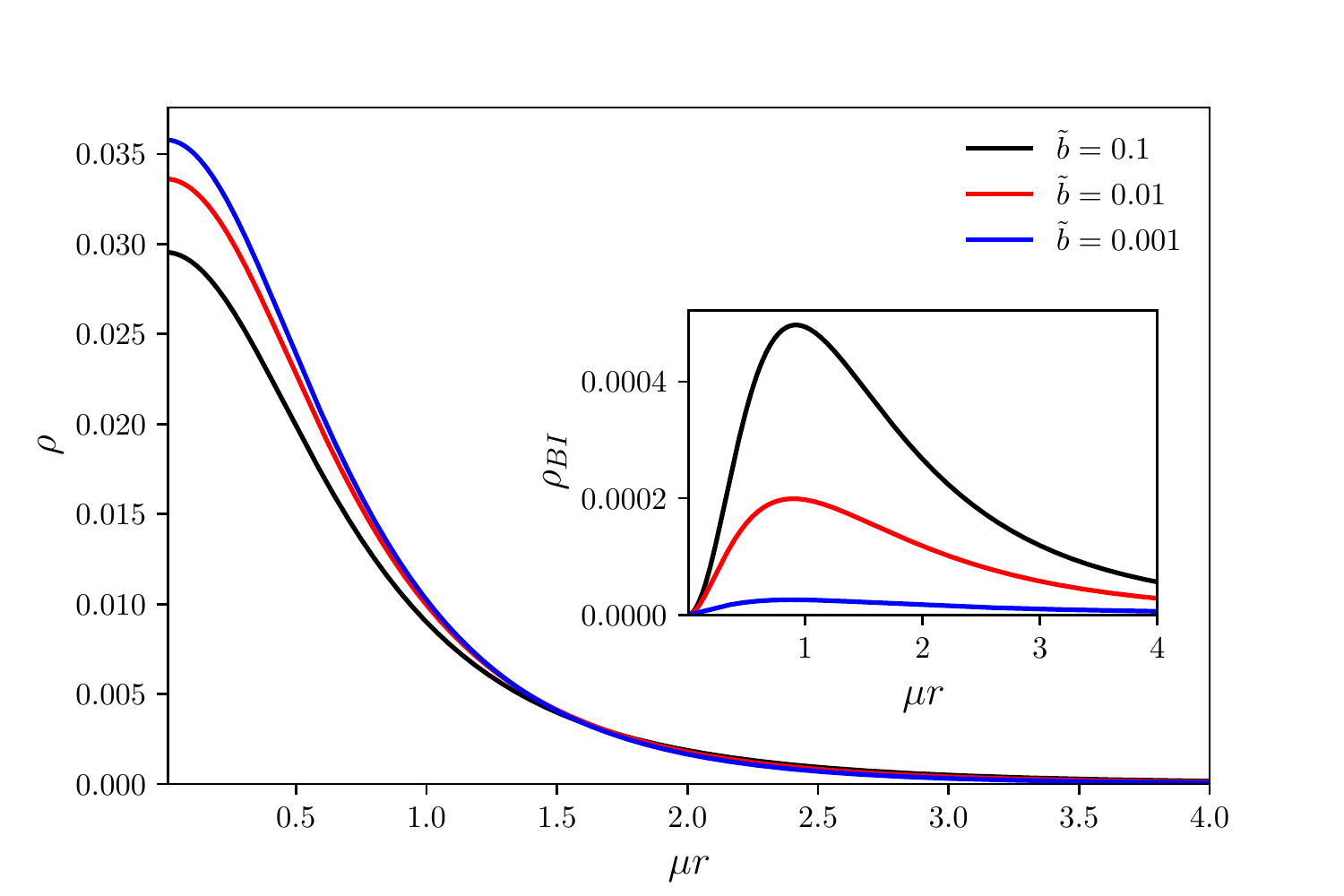}
  \caption{BI boson star solutions. Left panel: Scalar field, electromagnetic potential, lapse and conformal factor for a solution with $\tilde{q}=0.65$, $\tilde{b}=0.01$ and $\tilde{\phi}_{0}=0.5$. Right panel: Energy density (\ref{eq:density}) for solutions with $\tilde{q}=0.65$, $\tilde{\phi}_0=0.5$ and three different values of the BI parameter $b$.}
  \label{fig:single_solution}
\end{figure}
Fig.~\ref{fig:single_solution} shows the four fields $\{\tilde \phi,C,N,\Psi\}$ for a single solution with $(\tilde{q},\tilde{b},\tilde{\phi_0})=(0.65,0.01,0.5)$ and the energy density as given in Eq. \eqref{eq:density}.

\begin{figure}[ht]
\includegraphics[width=.45\linewidth]{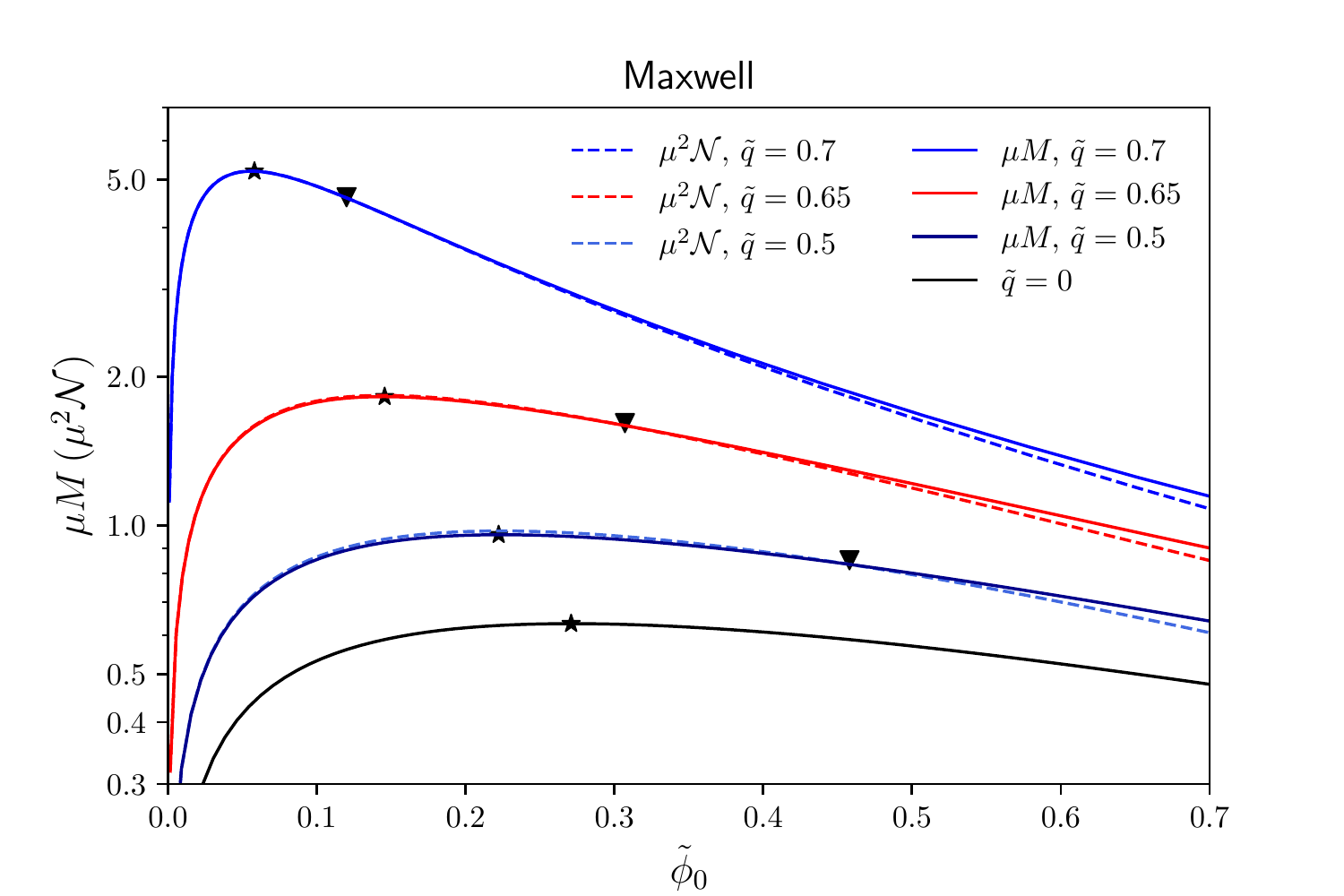}
\includegraphics[width=.45\linewidth]{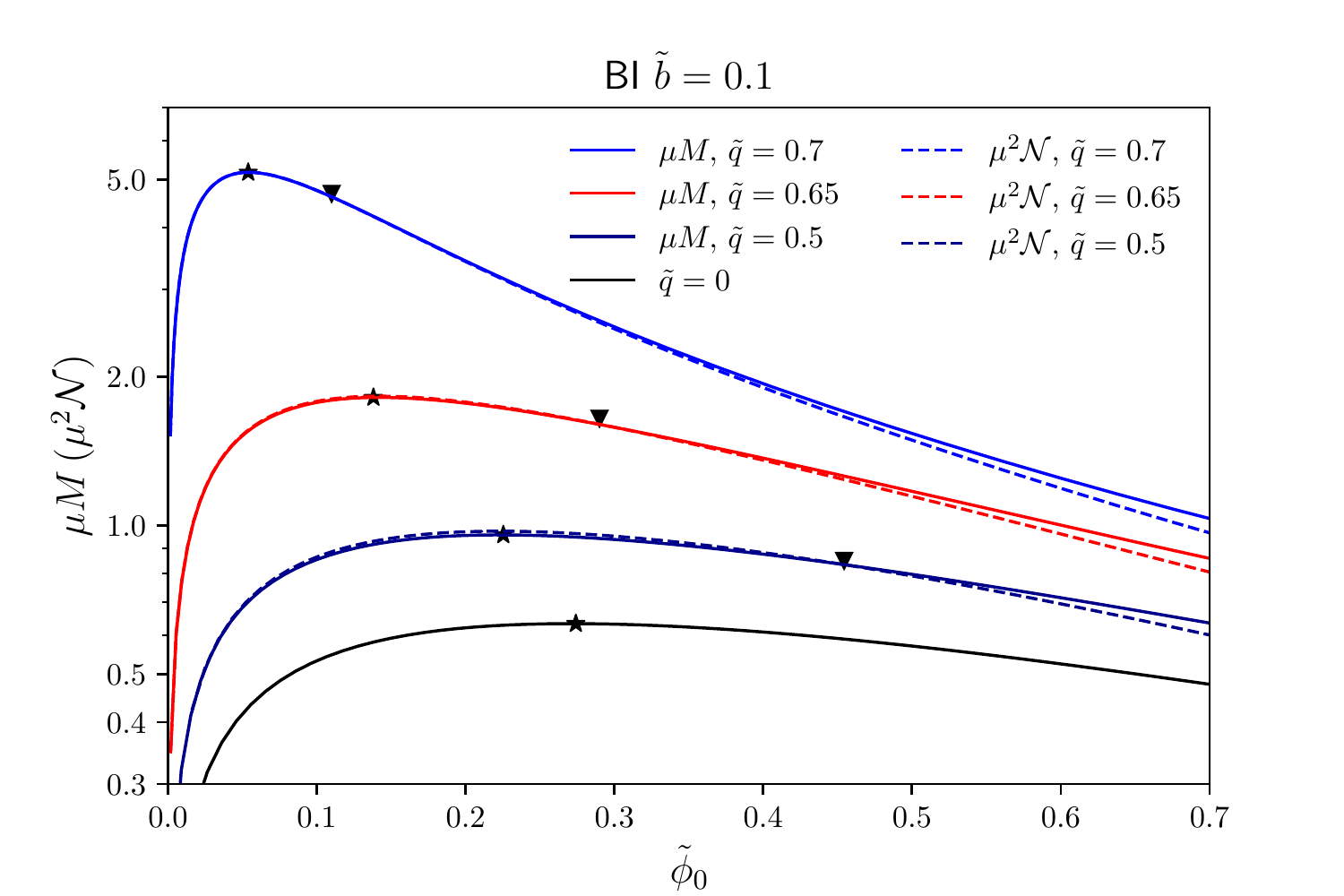}
\includegraphics[width=.45\linewidth]{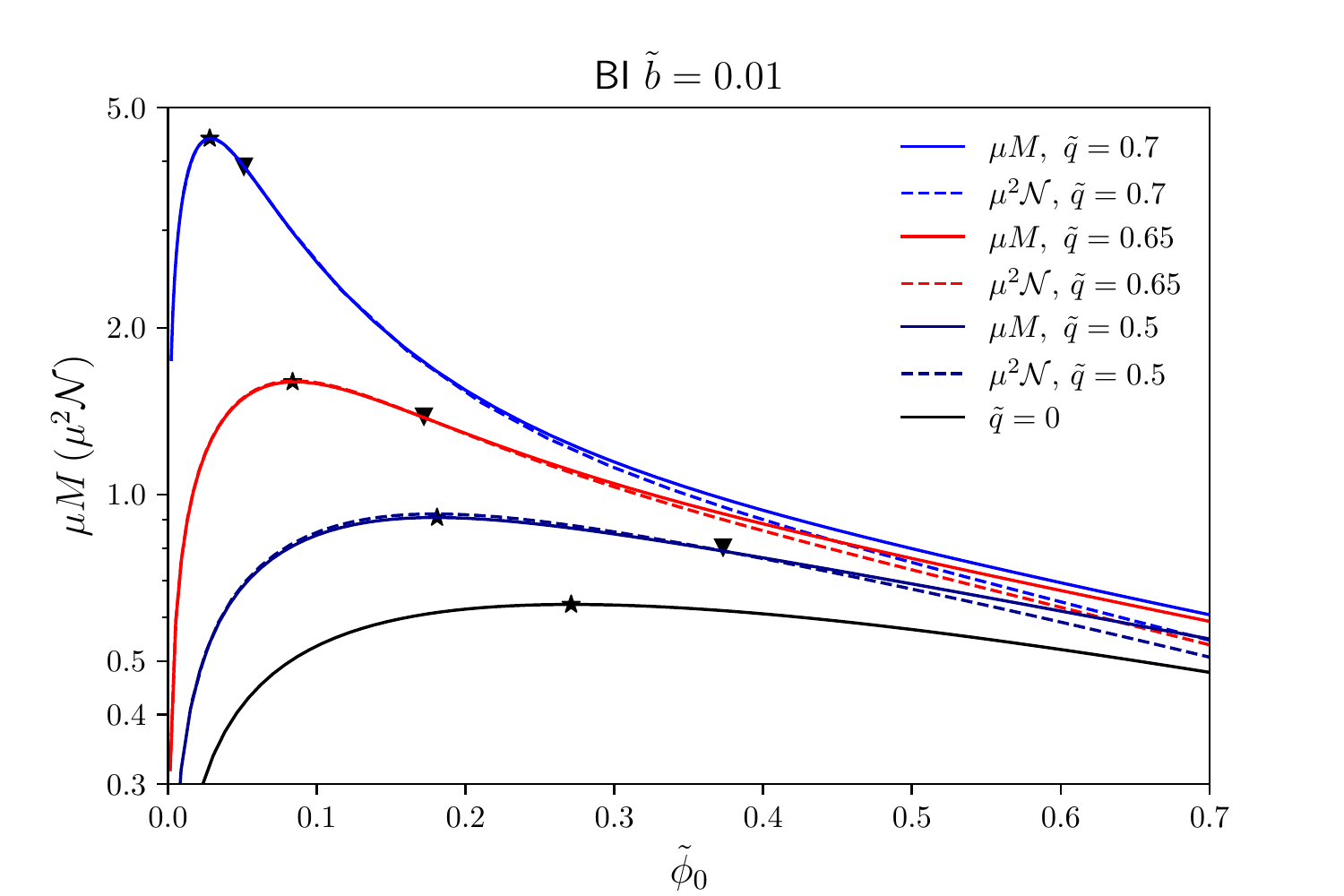}
\includegraphics[width=.45\linewidth]{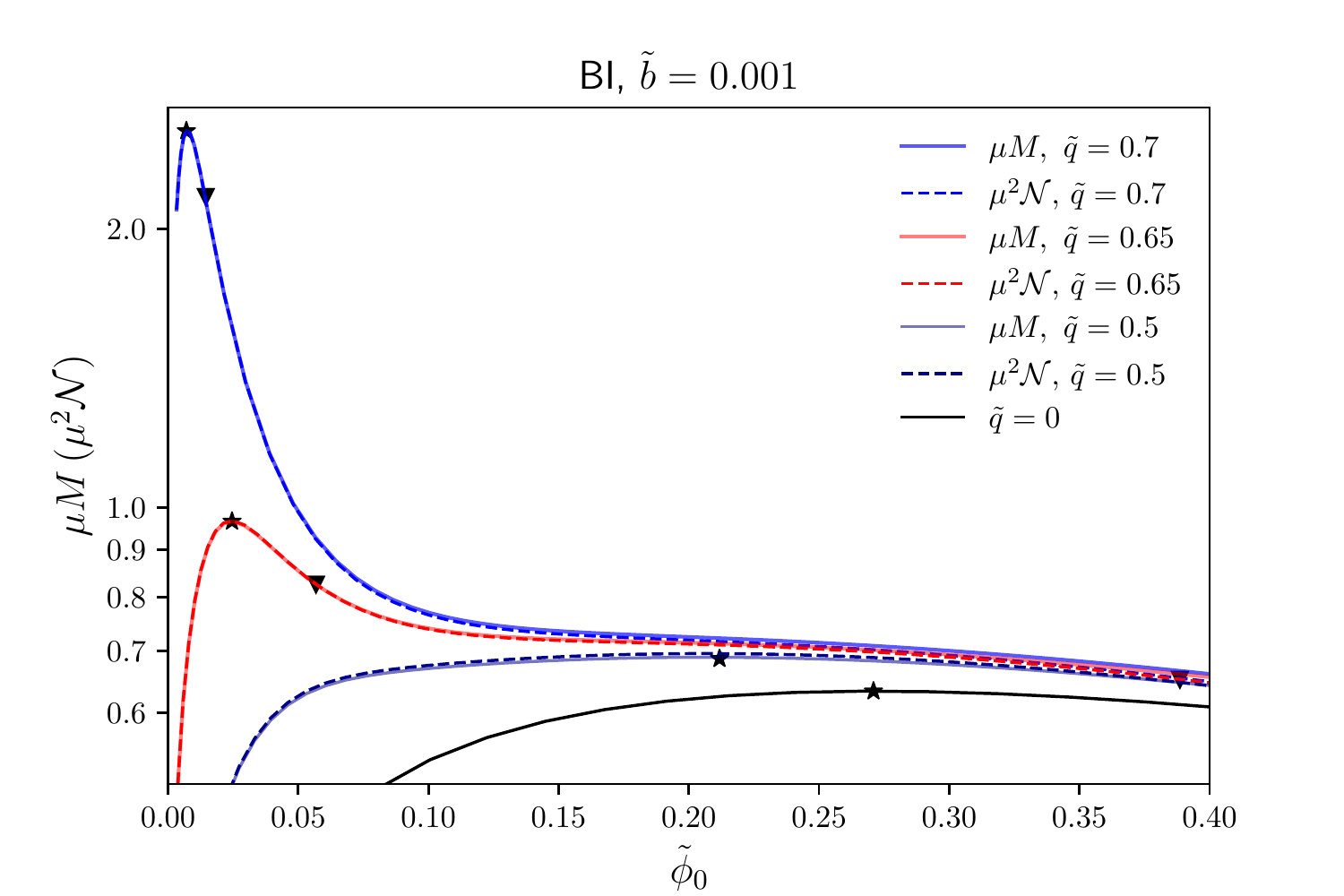}
 \caption{
   Mass and number of particles of charged boson stars in BI electrodynamics as a function of the central value of the scalar field for some representative values of $\tilde q$. The star symbol correspond to the maximum mass $M_{\mathrm{max}}$ and the inverted triangle marks the configuration with zero binding energy.
   } 
  \label{fig:Mass_vs_omega_qs}
\end{figure}
\begin{figure}
  \includegraphics[width=.45\linewidth]{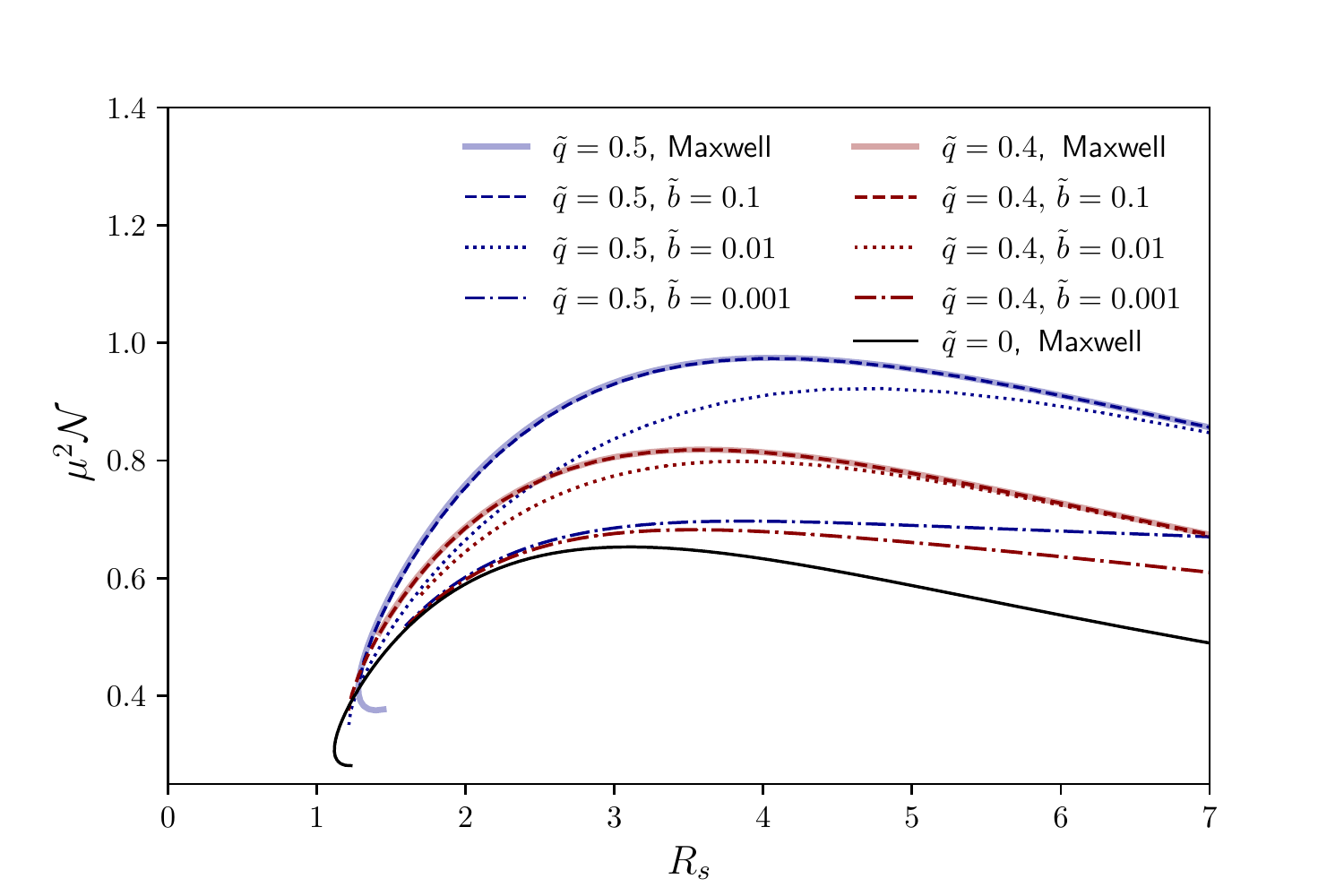}
  \caption{The particle number $\mathcal{N}$ as defined in \eqref{Npart} as a function of the effective radius of the star $R_s$.}
  \label{fig:N_vs_R99}
\end{figure}

The mass $M$ corresponding to several configurations is plotted in Fig.~\ref{fig:Mass_vs_omega_qs} as functions of the central value $\tilde \phi_0$, for different values of the boson charge $q$ and different values of $b$. For the Maxwell's electrodynamics the behaviour of the total mass $M$ as function of $\tilde \phi_0$ has been reported for instance in Ref.~\cite{Jetzer:1989av}. 
For a given value of $\tilde \phi_0$ the mass of the configurations increases as the charge increases.
For values $\tilde{b}>1$ the behaviour is qualitatively similar to Maxwell but for values $\tilde{b}<1$ the behaviour of $M$ changes, namely, for the same value of $\tilde q$ and $\tilde \phi_0$ the mass of the BI case is slightly lower than the Maxwell counterpart.
Furthermore, the maximum mass configuration $M_{\mathrm{max}}$, is attained with a lower value of $\tilde \phi_0$ for a fixed value of $\tilde q$. A quite similar behaviour is displayed by the number of particles also plotted in Fig.~\ref{fig:Mass_vs_omega_qs}. 
A linear stability analysis of charged boson stars in Maxwell electrodynamics performed by Jetzer \cite{Jetzer:1989us} shows that configurations on the right of the maximum mass configuration are unstable under small perturbations (see the plot on the left in the first row of Fig.~\ref{fig:Mass_vs_omega_qs} ) but configurations on the left of the maximum are stable. We expect that the introduction of BI electrodynamics will not have a significant effect on the stability properties of the charged stars but a thorough analysis has to be performed. 
The number of particles as a function of the effective size of the star
$R_s$, is displayed in Fig.~\ref{fig:N_vs_R99}. 
As in Maxwell boson stars, the number of particles in BI boson stars increases with the effective size of the star up to a maximum value and then starts to decrease for larger radii. This behaviour is qualitatively similar as the charge $\tilde q$ increases but the value of the maximum mass increases with $\tilde q$.
Since configurations with larger radius are less massive this yields to less compact stars as shown below.
%

\begin{figure}
  \includegraphics[width=.45\linewidth]{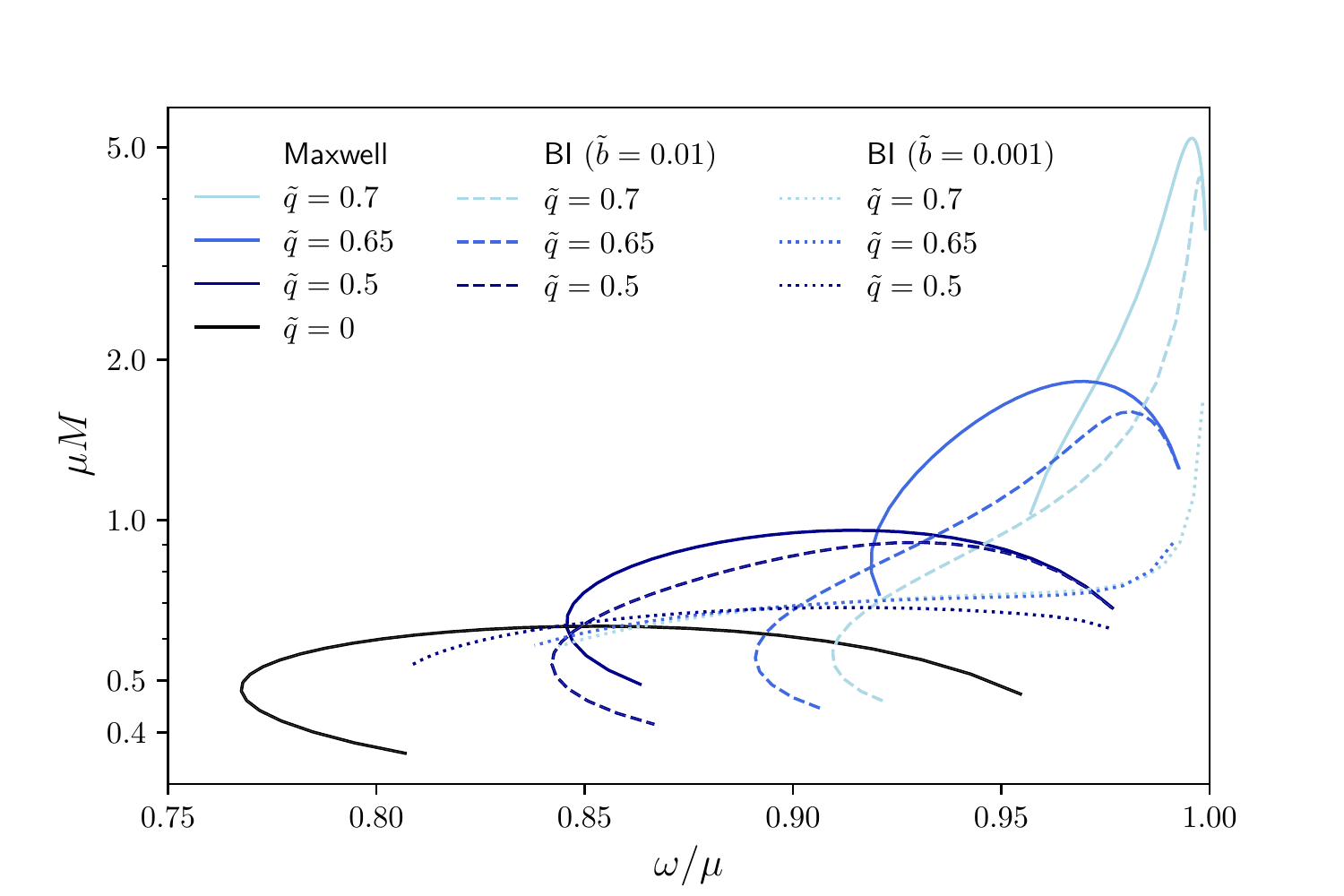}
  \caption{Space of solutions for BI and  Maxwell's boson stars. The mass of spherical BI boson stars (dashed) are shown for some values of the parameter $b$. In the limit $b\rightarrow\infty$ we recover the solutions for Maxwell boson stars (solid). Although the profiles are qualitatively similar for small values of $b$, the maximum mass of the BI boson stars is smaller than Maxwell's for any value of $b$.}
  \label{fig:mass_vs_w_qs}
\end{figure}

Fig.~\ref{fig:mass_vs_w_qs} shows a typical inspiral curve of the mass, in a space of solutions, as a function of the frequency for some values of $\tilde q$. As mentioned above, generically, the effect of the charge is to increase the value of $M$, furthermore, one can find configurations with higher values of mass and a smaller range of $\omega$ for larger values of $q$. 
The difference between Maxwell and BI configurations is that configurations with the same frequency have a lower mass in BI. This difference however, vanishes for values $\omega \sim \mu$. 

\begin{figure}
  \includegraphics[width=.45\linewidth]{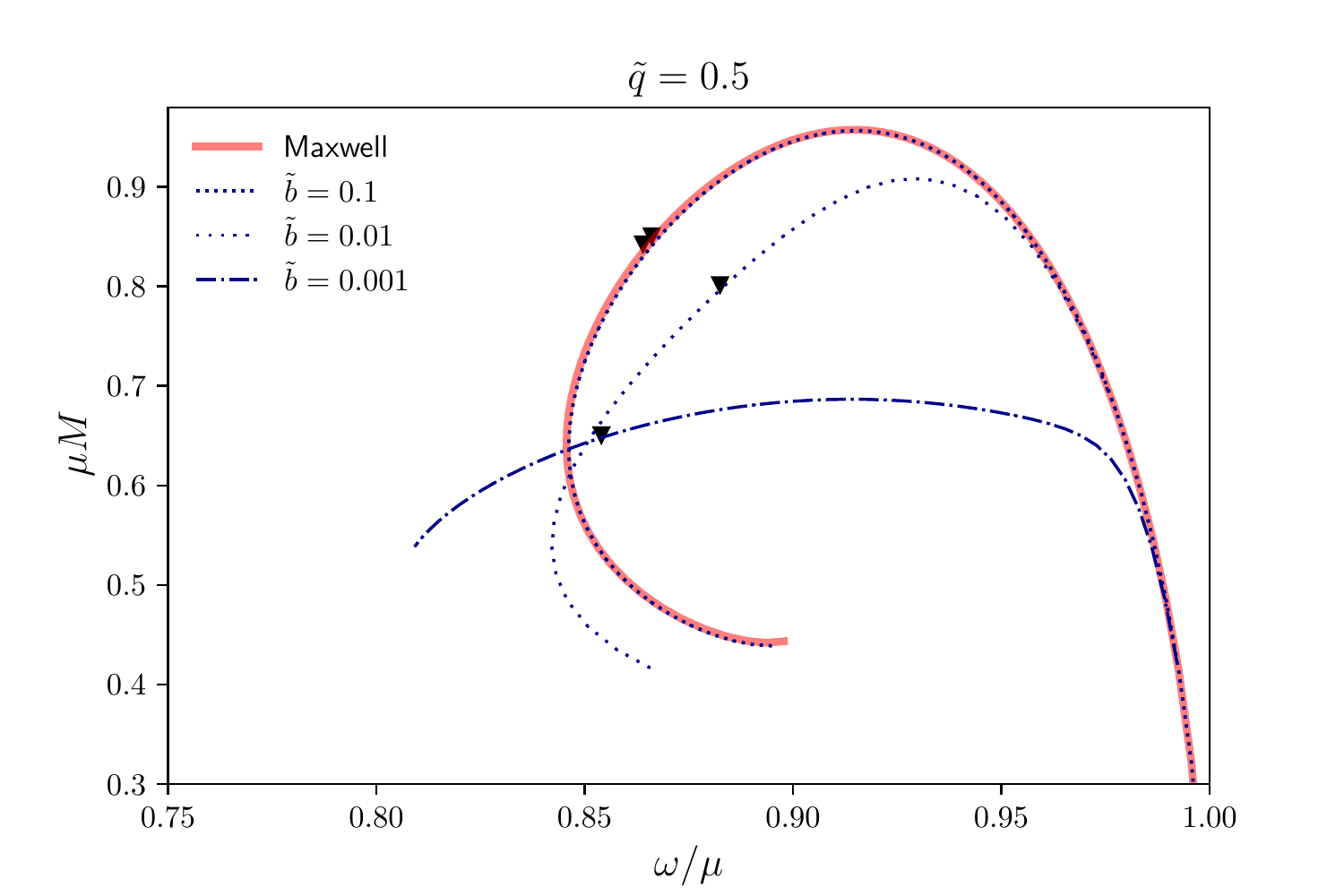}
  \caption{Space of solutions of boson stars with a fixed charge $q$. As the value of $b$ decreases the
   mass of the stars decreases for the same frequency. The inverted triangles represent boson stars with zero binding energy.
    }
  \label{fig:mass_vs_w_bs}
\end{figure}

Fig.~\ref{fig:mass_vs_w_bs} displays the behaviour of the mass of charged configurations as a function of $\omega$ with a (representative) fixed charge $\tilde q=0.5$ for some values of $\tilde b$. As the value of $\tilde b$ decreases the mass also decreases for stars with the same frequency $\omega$.

\begin{figure}
 \includegraphics[width=0.45\textwidth]{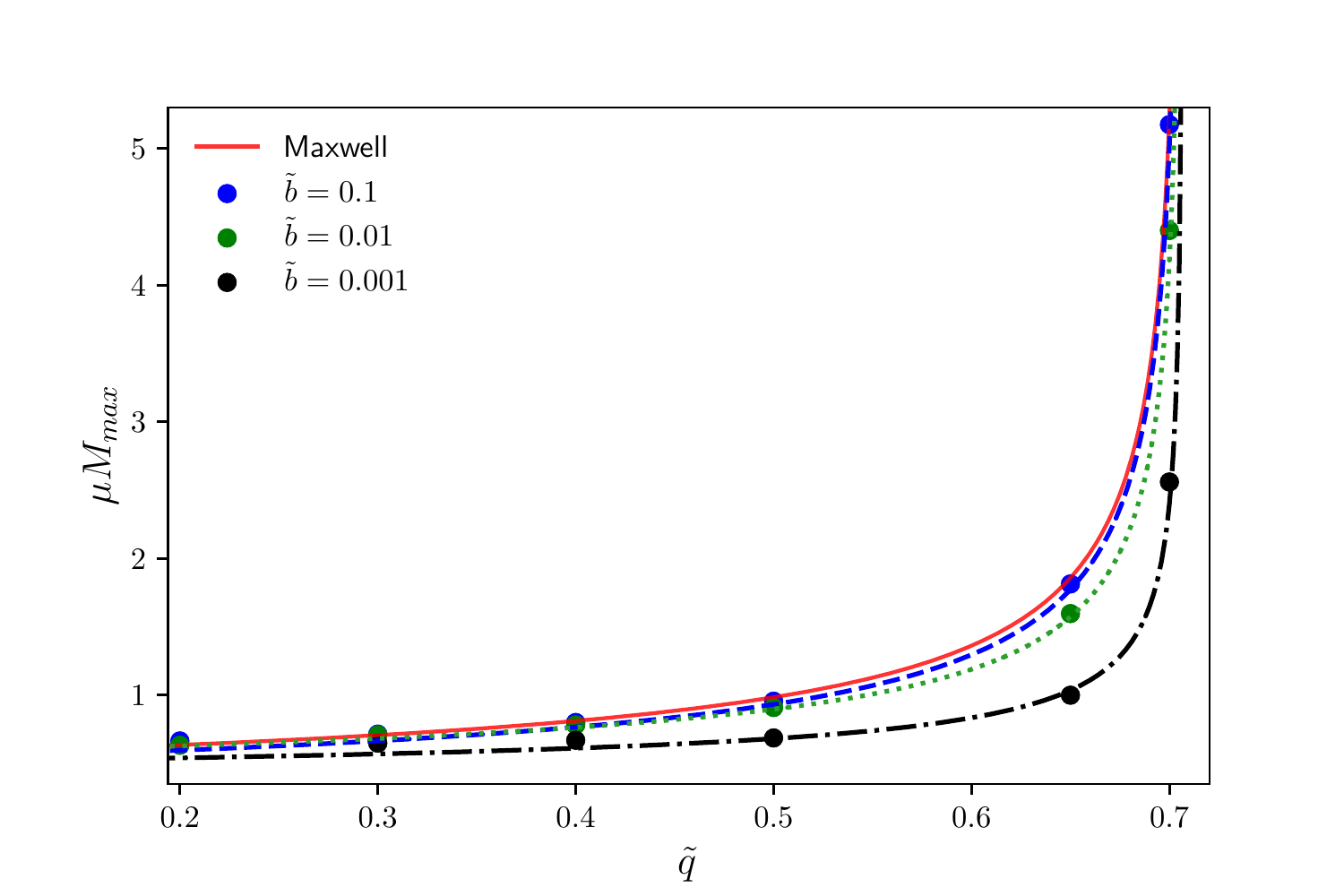}
 \caption{
 Maximum possible mass of charged boson stars as a function of the charge $q$ for fundamental (with zero nodes) configurations. Solutions with a fixed $q$ in Maxwell electrodynamics have a larger mass as compared to BI. 
 The behavior of the mass can be fit with a function of the form $M_{\mathrm{max}}\approx\alpha_{N}\left(q_{ref}-q\right)^{-1/2}+\beta_{N}$. The dots correspond to the numerically calculated solutions. The values of the parameters are given in Table~\ref{Tab}.
}
\label{fig:Mvsq} 
\end{figure}

As stated before, the maximum mass of the stars $M_{\mathrm{max}}$, grows with the charge for both Maxwell and BI electrodynamics, however the growth rate is larger in the Maxwell case. Fig.~\ref{fig:Mvsq} displays the maximum mass attained by the stars as a function of $\tilde q$ for some values of $\tilde b$. One can see that a larger charge is needed, as the BI parameter decreases, to get the same maximum mass than the Maxwell electrodynamics. Moreover the maximum possible mass of a charged boson star with fixed $\tilde q$ is obtained with a Maxwell field, independently of the central value of the field $\tilde \phi_0$.
In order to determine the grow rate of the maximum mass as a function of the charge, we use a fit of the form $M_{\mathrm{max}}\approx\alpha_{N}\left(q_{ref}-q\right)^{-1/2}+\beta_{N}$. 
The values of the constants $\alpha_N$, $\beta_N$ and $q_{ref}$ corresponding to configurations with $\tilde{b} = 0.1, 0.01$ and $0.001$ are given in Table \ref{Tab}.
\begin{table}[ht]

\caption{The maximum mass attained by the BI boson stars $M_{\mathrm{max}}$, as a function of the charge of the field can be fit with a function of the form
$M_{\mathrm{max}}\approx\alpha_{N}\left(q_{ref}-q\right)^{-1/2}+\beta_{N}$. The coefficients were found with a least square algorithm.}
\centering{}%

\begin{tabular}{ccccccc}
\hline 
$\Tilde{b}$&  & $q_{ref}$& &$\alpha_{N}$& &$\beta_{N}$  \tabularnewline
\hline 
\hline 
$\infty$ &  &$0.71$ &  & $4.37\times10^{-1}$ &  & $0.11\times10^{-1}$ \tabularnewline
$0.1$ &  & $0.71$ &  & $4.33\times10^{-1}$ &  & $0.22\times10^{-1}$ 
\tabularnewline
$0.01$ &  & $0.71$ &  & $3.60\times10^{-1}$ &  & $1.25\times10^{-1}$ \tabularnewline
$0.001$&  & $0.71$ &  & $1.88\times10^{-1}$ &  & $2.90\times10^{-1}$  \tabularnewline
\hline 
\end{tabular}\label{Tab}
\end{table}

\begin{figure}
 \includegraphics[width=0.45\textwidth]{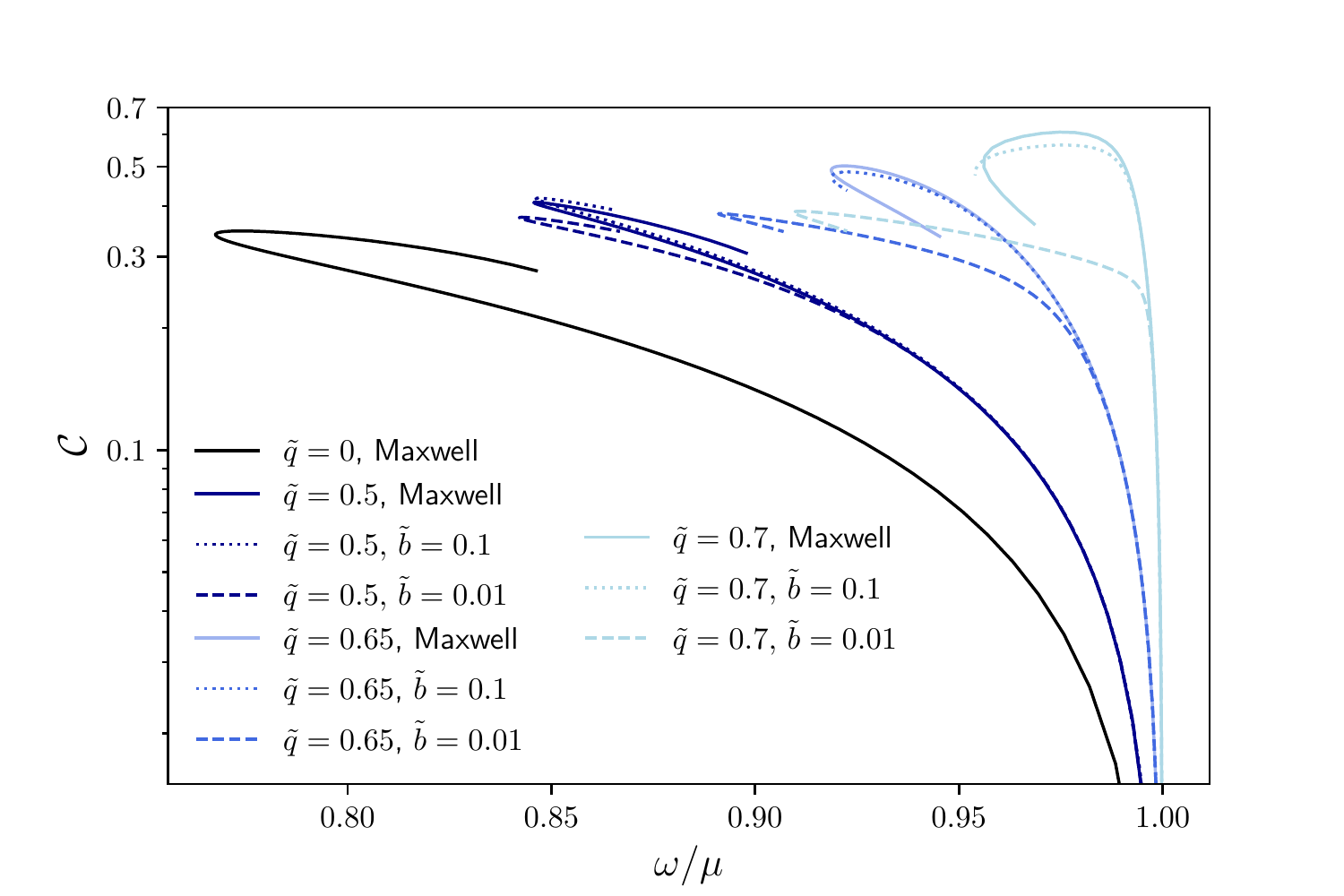}
 \caption{
Compactness of the charged boson stars as a function of the frequency for some  values of $q$ for both BI and Maxwell electrodynamics. 
}
\label{fig:C99omega} 
\end{figure}

One of the main effects of the charge in boson stars is to increase their compactness \cite{Jetzer:1989av}. Configurations with large charge $q$ tend to be more compact. For BI electrodynamics the same trend follows. The most important effect is that stars with lower frequency exist for small values of $b$.
For the non charged case the maximum compactness of a star ${\mathcal{C}_{\mathrm{max}}}=\frac{M_{\mathrm{max}}}{R_s}$, is of the order of ${\mathcal{C}}^{q=0}_{\mathrm{max}}\sim 0.347$. Higher values of $\tilde q$ yield configurations with higher values of the maximum compactness, for instance ${\mathcal{C}}^{q=0.7}_{\mathrm{max}}\sim 0.609$. The definition used for $\mathcal{C}$ make use of the quantity $R_s$ which we have seen to be significantly smaller than the $R_{99}$ radius, normally used for boson stars. Since the mass used in the definition of compactness is the total mass, the compactness can exceed the value 0.5. By obtaining the $C_{99}:=M/R_{99}$ compactness of the mentioned configurations we have verified that charged configurations in both Maxwell and BI electrodynamics are quite far from the Buchdahl limit for stable configurations \cite{Buchdahl:1959zz}.
The compactness of some solutions for both Maxwell and BI are shown in Fig. \ref{fig:C99omega}. 
%
\subsection{Binding energy}

\begin{figure}[ht]
\includegraphics[width=.45\linewidth]{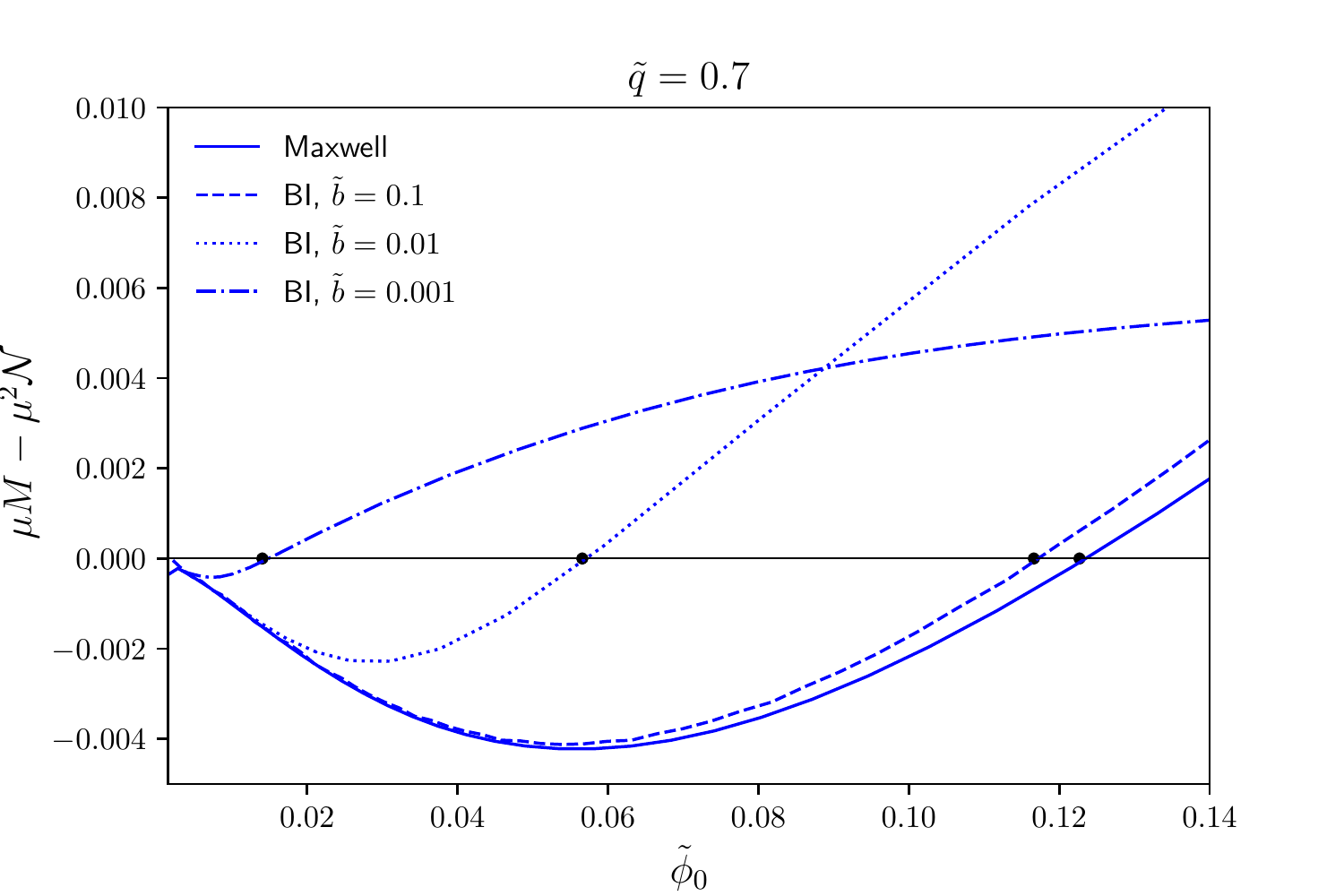}
\includegraphics[width=.45\linewidth]{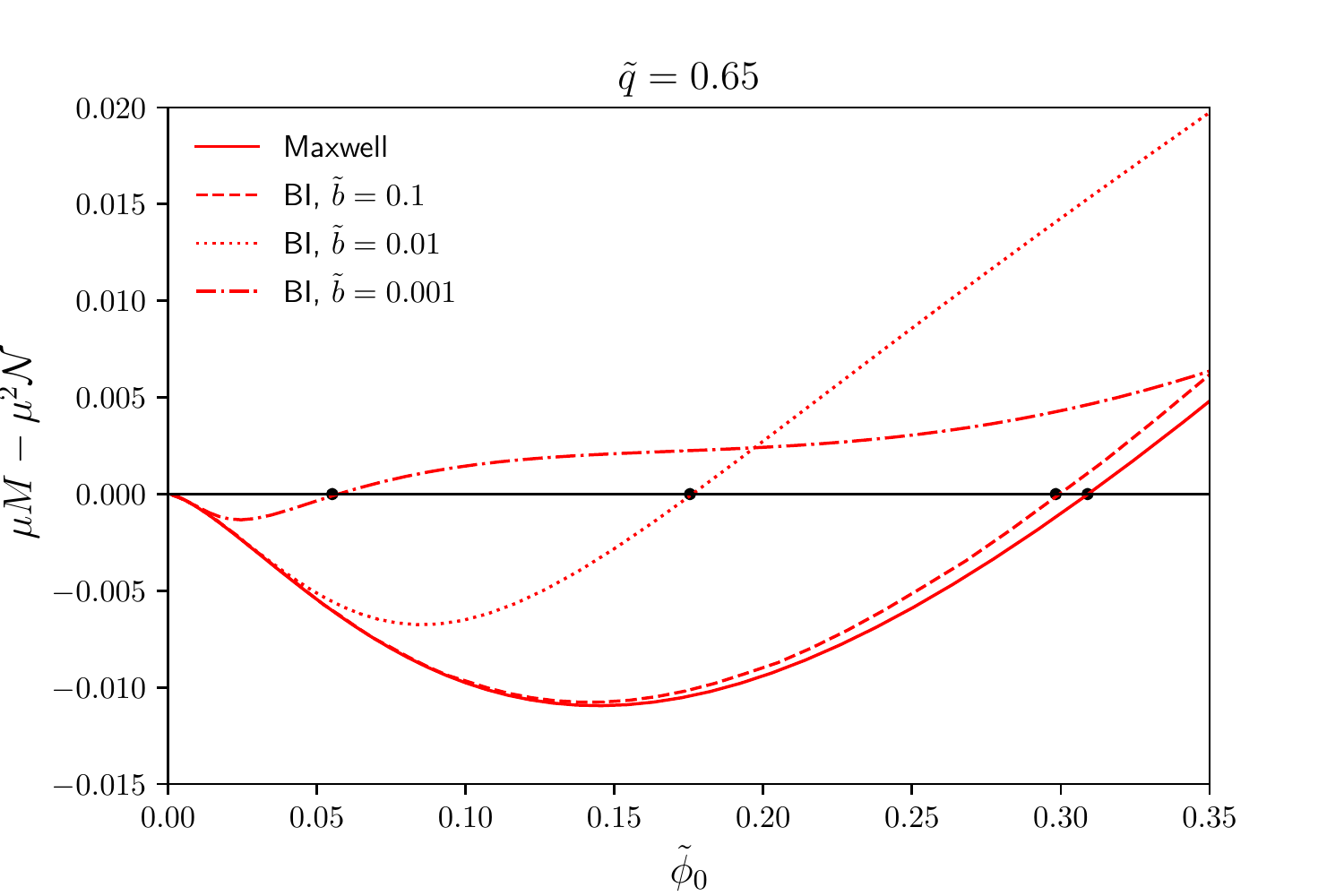}
\includegraphics[width=.45\linewidth]{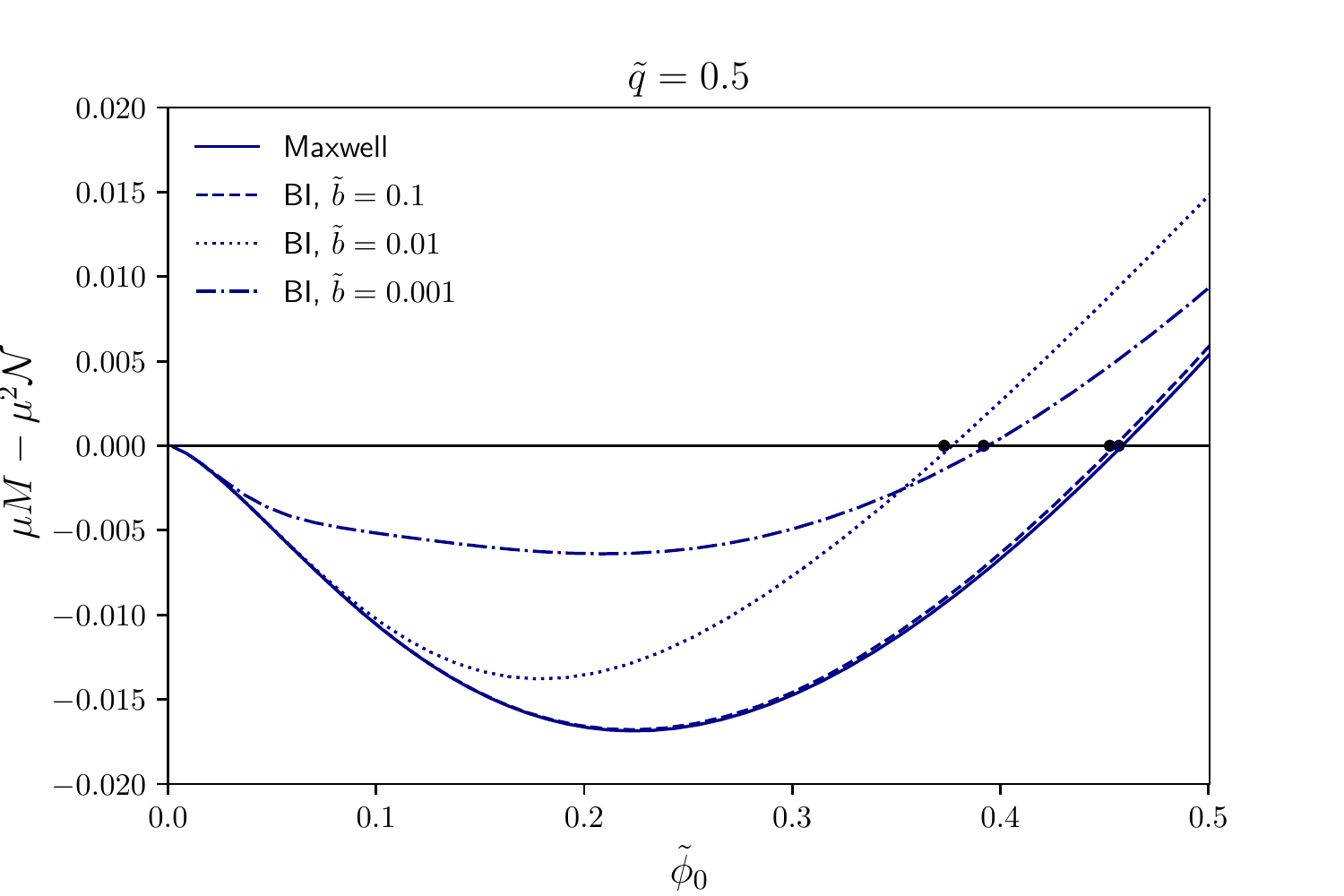}
  \caption{Binding energy as defined in \eqref{eq:binding-e}. The transition configurations are marked with a black dot. As the value of $q$ increases, less configurations with with negative binding energy are found.}
  \label{fig:N-M}
\end{figure}

The gravitational binding energy of charged boson stars is determined as the difference between the total mass and the number or particles times the mass of each particle 
\begin{equation}\label{eq:binding-e}
    U_{\rm bin} = M - \mu \mathcal{N} \ . 
\end{equation}
This quantity has been used to characterize non-charged boson stars according to its final state after some perturbation were induced \cite{Seidel:1990jh}.
For charged boson stars, the binding energy can take positive or negative values depending on the values of $\phi_0$ and $\tilde q$. 
Fig.~\ref{fig:N-M} shows the binding energy as a function of $\phi_0$ for some values of $\tilde q$ and representative values of $\tilde b$. As a reference, the Maxwell case is also displayed. 

Configurations in which the binding energy transit from negative to positive are in the right of the maximum mass configuration in a plot of mass vs $\phi_0$ as displayed in Fig. \ref{fig:Mass_vs_omega_qs}.  
As the value of the charge $\tilde q$ increases, less configurations with negative binding energy exist due to the electric repulsion. However, as $\tilde b$ decreases, almost all BI boson stars have positive binding energy for larger values of $\tilde q$.  
In Table \ref{tab:Mmax} we present 
the maximum mass $M_{\mathrm{max}}$ attained by charged boson stars for some values of $\tilde q$ and $\tilde b$. Additionally, configurations with zero binding energy are also displayed. This sample of solutions represent the main differences between Maxwell and BI boson stars.
\begin{center}
\begin{table}
\centering{} \caption{ 
Maximum mass of boson stars $M_{\mathrm{max}}$ for a given value of the charge $\tilde q$. Configurations with $\mu M=\mu^{2}\mathcal{N}$ have zero binding energy and $\tilde \phi_0$ corresponds to the central value of the scalar field. All the quantities are given in terms of the mass of the scalar field $\mu$. The corresponding scaling is given in Eq.~\eqref{eq:tilde}.
} \begin{tabular}{c||ccc|cccc||ccccc||ccc|ccc}
\hline 
$\tilde{q}_{(b\rightarrow\infty)}$ &  & $\mu M_{\mathrm{max}}$ &  & $\mu M=\mu^{2}\mathcal{N}$ &  & $\tilde{\phi}_{0}$ &  &  & $\tilde{b}$ &  & $\tilde{q}$ &  &  & $\mu M_{\mathrm{max}}$ &  & $\mu M=\mu^{2}\mathcal{N}$ &  & $\tilde{\phi}_{0}$\tabularnewline
\hline 
\hline 
$0.7$ &  & $5.204$ &  & $5.202$ &  & $0.123$ &  &  & $0.1$ &  & $0.7$ &  &  & $5.175$ &  & $4.675$ &  & $0.110$\tabularnewline
$0.65$ &  & $1.822$ &  & $1.821$ &  & $0.309$ &  &  & $0.01$ &  & $0.7$ &  &  & $4.398$ &  & $3.905$ &  & $0.051$\tabularnewline
$0.60$ &  & $1.326$ &  & $1.160$ &  & $0.381$ &  &  & $0.001$ &  & $0.7$ &  &  & $2.556$ &  & $2.165$ &  & $0.015$\tabularnewline
$0.59$ &  & $1.268$ &  & $1.107$ &  & $0.392$ &  &  & $0.1$ &  & $0.65$ &  &  & $1.815$ &  & $1.641$ &  & $0.298$\tabularnewline
$0.58$ &  & $1.217$ &  & $1.065$ &  & $0.402$ &  &  & $0.01$ &  & $0.65$ &  &  & $1.597$ &  & $1.381$ &  & $0.175$\tabularnewline
$0.5$ &  & $0.958$ &  & $0.836$ &  & $0.457$ &  &  & $0.001$ &  & $0.65$ &  &  & $0.966$ &  & $0.824$ &  & $0.055$\tabularnewline
$0.4$ &  & $0.801$ &  & $0.696$ &  & $0.495$ &  &  & $0.1$ &  & $0.5$ &  &  & $0.957$ &  & $0.485$ &  & $0.117$\tabularnewline
$0.1$ &  & $0.640$ &  & $0.561$ &  & $0.536$ &  &  & $0.01$ &  & $0.5$ &  &  & $0.909$ &  & $0.482$ &  & $0.057$\tabularnewline
 &  &  &  &  &  &  &  &  & $0.001$ &  & $0.5$ &  &  & $0.686$ &  & $0.541$ &  & $0.014$\tabularnewline
\hline 
\end{tabular}\label{tab:Mmax}
\end{table}
\par\end{center}


\subsection{Self Interaction}

The effect of a quartic self-interaction in BI boson stars is quite similar as the effect described in \cite{Jetzer:1989av} for charged  boson stars in Maxwell electrodynamics. That is, the self-interaction in the scalar field allows to get more massive configurations than the non-interacting case. Also, 
as the value of the self coupling constant  $\tilde \lambda$ increases, the mass of the stars increases for a given (fixed) value of  
$\tilde \phi_0$. This effect however, is more noticeable for larger values of $\tilde \phi_0$. 

\begin{figure}[ht]
\includegraphics[width=.45\linewidth]{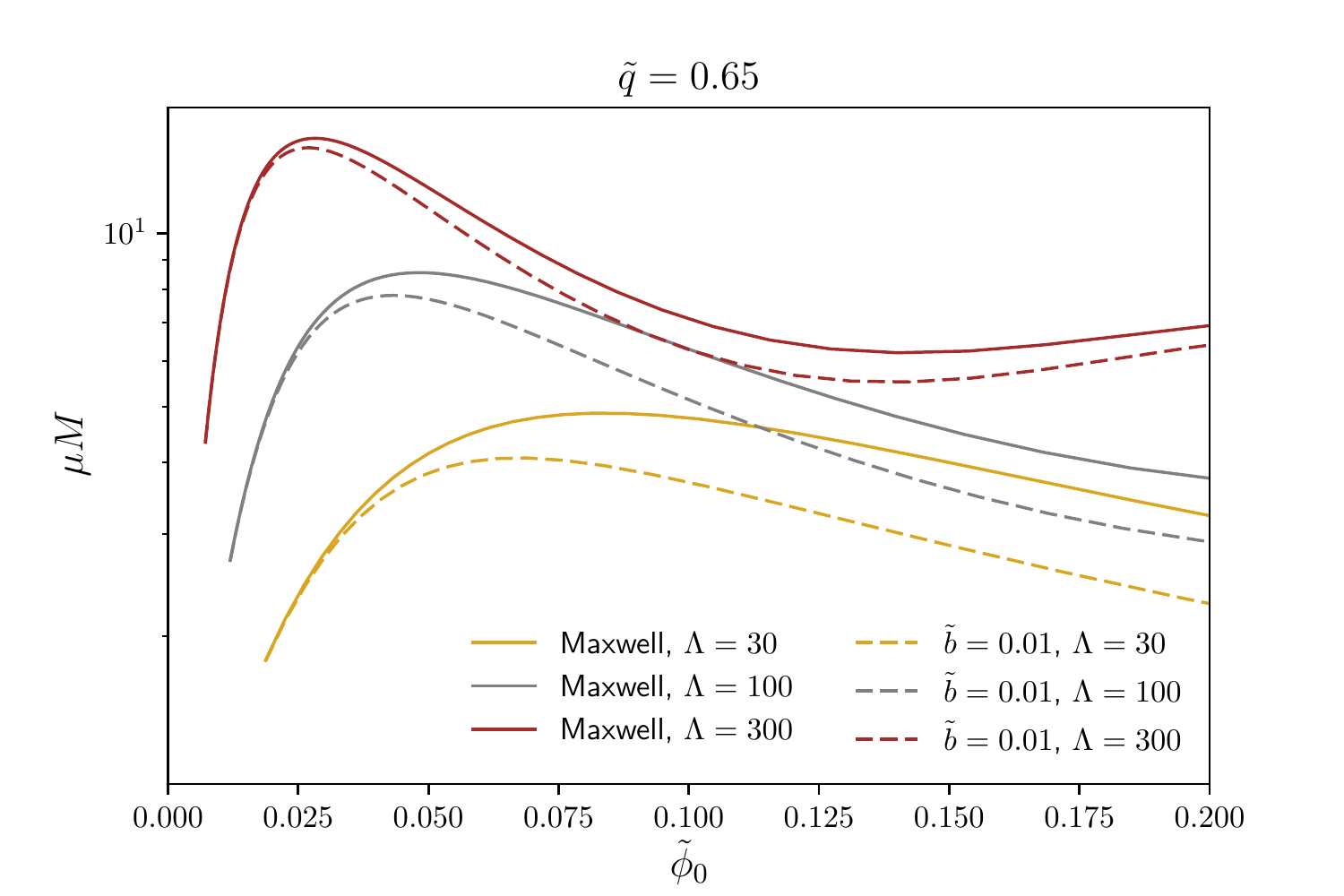}
 \caption{
   Mass of charged boson stars as a function of the central value of the scalar field for $\tilde{q}=0.65$ and three cases of the self-interaction parameter $\Lambda=4\pi \tilde{\lambda}$. The quantity $\Lambda$ coincides with the self-interaction parameter of both Colpi \textit{et al.} \cite{Colpi:1986ye} and Jetzer \cite{Jetzer:1989av}.
   }
  \label{fig:phiM-lambda}
\end{figure}

In Fig.~\ref{fig:phiM-lambda} we show the sequence of solutions for BI and Maxwell electrodynamics of equilibrium solutions with $\tilde{q}=0.65$ and three cases of self-interacting scalar fields. The effect of $\tilde \lambda$ in BI boson stars is to increase the compactness and particle number besides the mentioned increase in the mass. Additionally, the effects of $\tilde b$ on the solutions previously discussed for non interacting fields remain when $\tilde \lambda\neq 0$. Configurations with other values of $\tilde q$ present the same behaviour.

\subsection{Asymptotic spacetime}

According to Eq.~\eqref{eq:kg_eq} the scalar field decays exponentially as $r\rightarrow \infty$. It is thus expected that the spacetime tends asymptotically to the static electrovacuum solution of the Einstein-BI theory of an object of mass $M$ and charge $Q$ in spherical symmetry. Such solution was introduced by Hoffman \cite{PhysRev.47.877} and its properties have been extensively described ever since, (see  for instance ~\cite{1984NCimB..84...65G, plebanski1970lectures,Breton:2002td} and references therein).
In Schwarzschild-like coordinates Hoffman's solution can be written as: 
$ds^{2}=-\mathcal{G}\,dt^{2}+{\mathcal{G}}^{-1}d R{}^{2}+R^{2}d\Omega^{2}$, where 
\begin{equation}\label{eq:metricBI-BH}
\mathcal{G}=1-\frac{2M}{R}+\frac{2}{3}b^{2}R^{2}\left(1-\sqrt{1+\frac{Q^{2}}{4\pi\,b^{2}R^{4}}}\right)+\frac{Q^{2}}{3\pi\,R}\int_{R}^{\infty}\frac{d\sigma}{\sqrt{\sigma^{4}+Q^{2}/4\pi\,b^{2}}} \ .
\end{equation}

The last term is an elliptic integral of the first kind,
which in the literature can be found written either in
terms of the Legendre’s elliptic integral: $\mathcal{F}(\beta,\kappa):=\int_{\beta}^{\infty}\left(1-\kappa^{2}\sin^{2}\sigma\right)^{-1/2}d\sigma$, or in terms of the hyper-geometric function $_{2}F_{1}\left(a,b;c;x\right)$ as follows 

\begin{eqnarray*}
\int_{R}^{\infty}\frac{d\sigma}{\sqrt{\sigma^{4}+Q^{2}/4\pi\,b^{2}}} & = & \frac{1}{2}\sqrt{\frac{\sqrt{4\pi}b}{Q}}\mathcal{F}\left(\arccos\left(\frac{\sqrt{4\pi}\, bR^{2}/Q-1}{\sqrt{4\pi}\, bR^{2}/Q+1}\right),\,\frac{1}{\sqrt{2}}\right)\\
 & = & \frac{1}{R}\,{}_{2}F_{1}\left(\frac{1}{4},\,\frac{1}{2};\,\frac{5}{4};\,-\frac{Q^{2}}{4\pi\,b^{2}R^{4}}\right).
\end{eqnarray*}

\begin{figure}
 \includegraphics[width=0.45\textwidth]{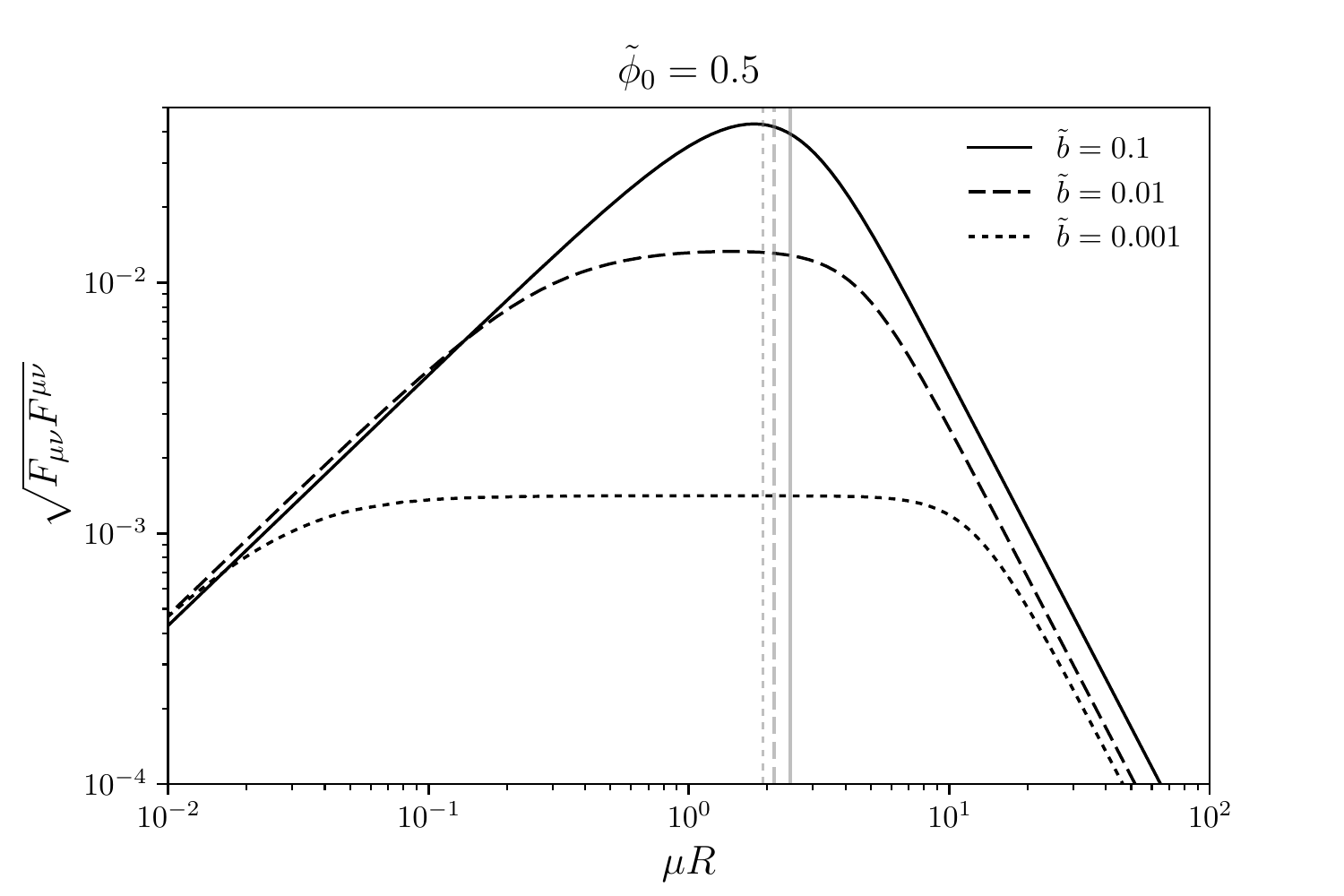} \includegraphics[width=0.45\textwidth]{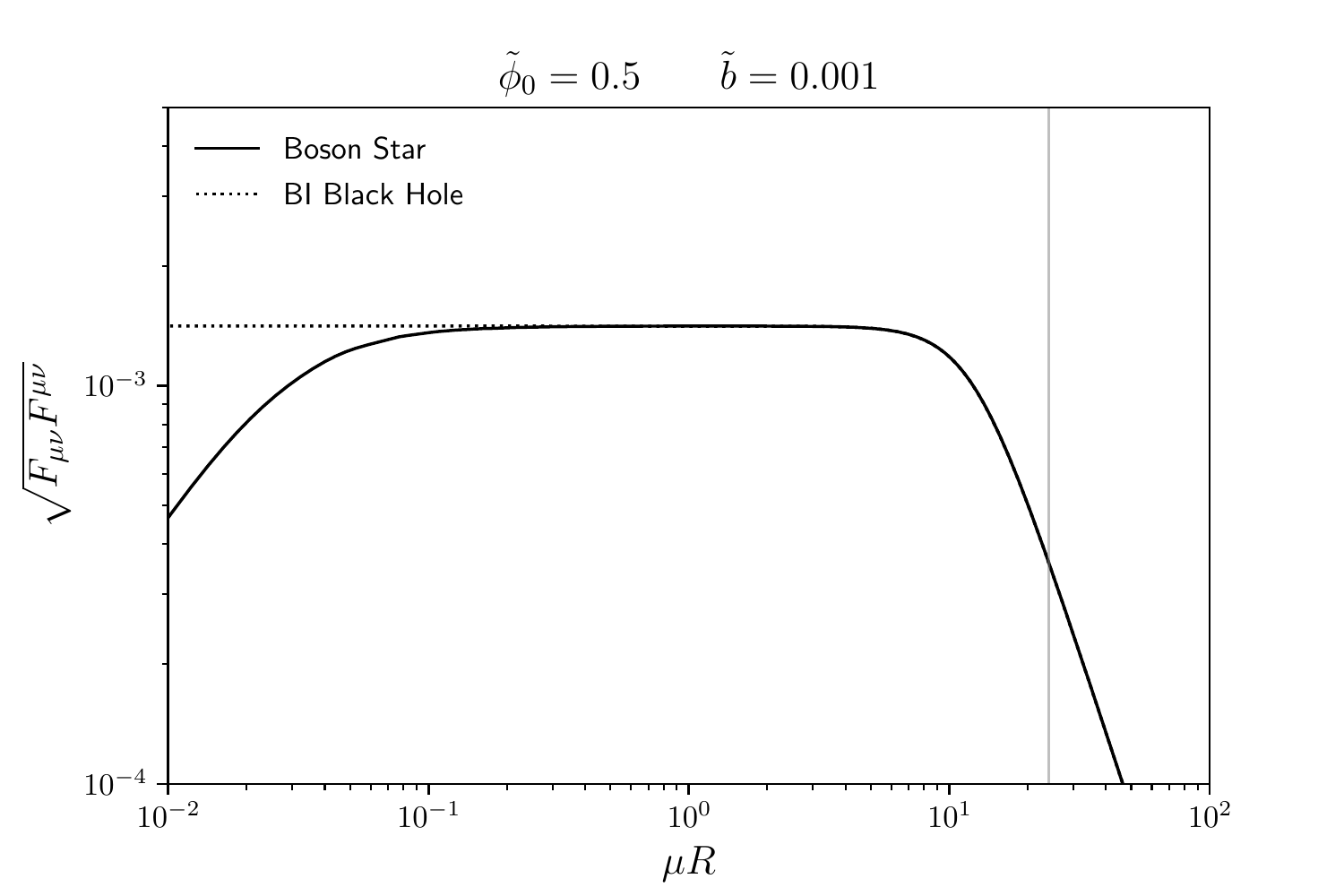}
 \caption{Invariant of the electromagnetic field (For BI boson stars $R=\Psi^2 r$ is the areal radius). Left panel: Three different solutions with the same charge $\tilde{q}=0.65$. Right panel: Comparison with the electrovacuum solution \cite{1984NCimB..84...65G} for the same values of $Q$ and $M$. 
 The radius $R_{s}$ for each star is indicated with vertical lines with values 2.4517, 2.1291 and 1.9262 for $\tilde{b}=$ 0.1, 0.01 and 0.001 respectively.
 }
\label{fig:invariantF} 
\end{figure}

Figure \ref{fig:invariantF} shows the electromagnetic invariant $F=F_{\mu\nu}F^{\mu\nu}$ for a boson star with $\tilde{\phi}_0 = 0.5$ and $\tilde{q}=0.65$ as a function of $r$ and $\tilde{b} = {0.1,0.01,0.001}$. Close to the origin this invariant (basically the square of the electric field in spherical symmetry) vanishes due to the conditions \eqref{eq:regularity_r} and grows with $r$ up to a maximum and then decreases asymptotically. 

Close to the origin it becomes more difficult to find boson stars solutions with small values of $\tilde b$ because the drastic change in the derivative of $F$ as $\tilde b$ decreases. This is consistent with the electromagnetic BI theory since large deviations of Maxwell electrodynamics may lead to nonphysical effects.

The asymptotic behaviour of $F$ for BI boson stars, on the other hand, is very similar to the one of Hoffman's solution with the same values of $M$ and $Q$. For $r\rightarrow \infty$ the invariant in both spacetimes is almost indistinguishable.
However, close to $r=0$ the electric field in Hoffman's solution has a constant nonvanishing value as shown the right panel of Fig. \ref{fig:invariantF}. 

\begin{figure}
\includegraphics[width=0.45\textwidth]{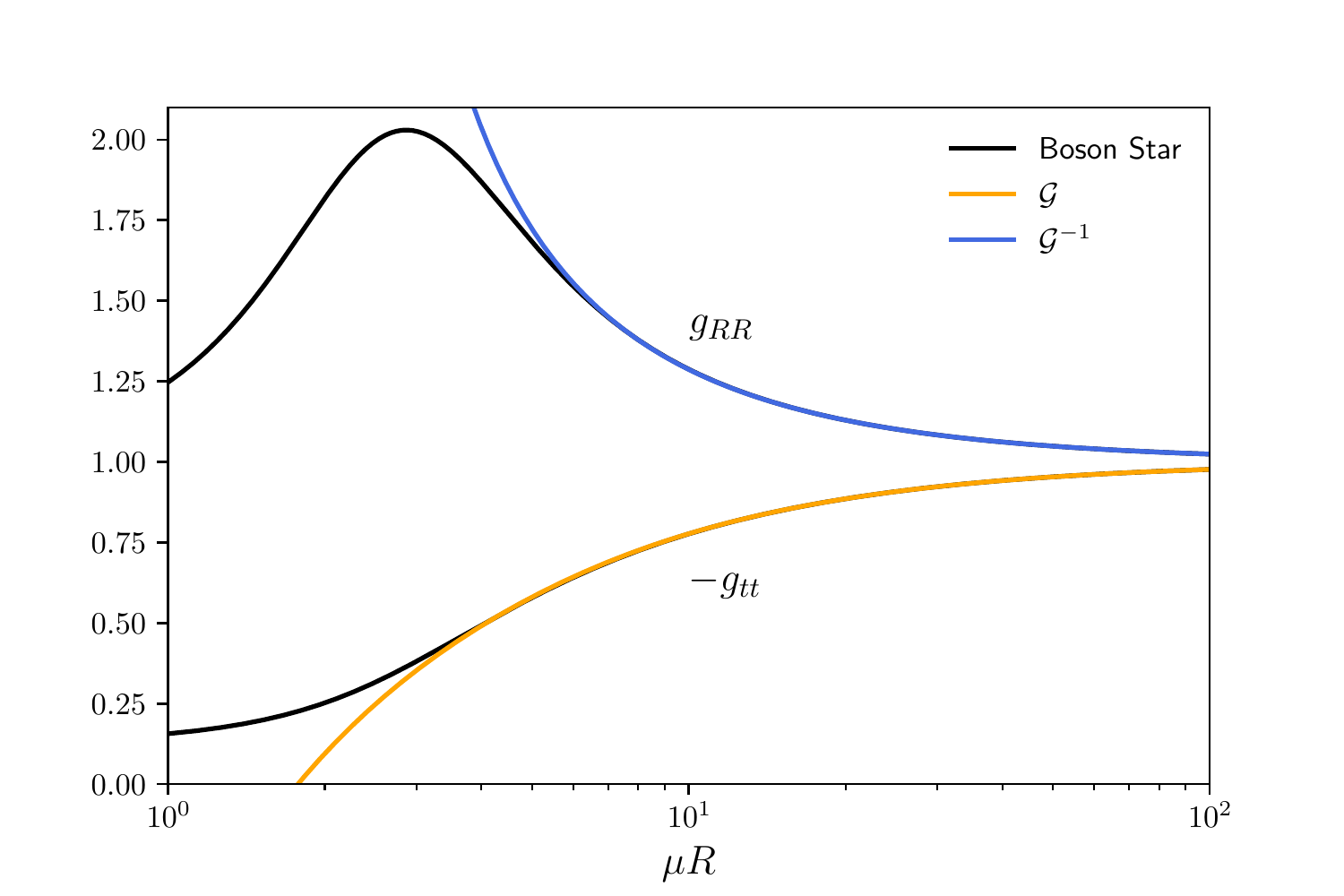}
  \caption{
  Metric components $g_{tt}$ and $g_{RR}$ (For BI boson stars $R=\Psi^2 r$ is the areal radius)  for a boson star with $(\tilde{q},\tilde{b},\tilde{\phi}_0)=(0.65,0.1,0.5)$ and the Hoffman's solution (\ref{eq:metricBI-BH}) with the same $M$ and $Q$. 
  Asymptotically, boson star spacetime matches the Hoffman's electrovacuum solution.}
  \label{fig:metric_vs_r}
\end{figure}

In Fig. \ref{fig:metric_vs_r} we show the metric components in Schwarzschild-like coordinates\footnote{The $g_{RR}$ component in $ds^2=g_{tt}dt^2+g_{RR}dR^2+R^2d\Omega^2$ is obtained from the numerical solutions of boson stars, which are in isotropic coordinates, using the transformation $g_{RR}=(1+2r\ d\ln\Psi/dr)^{-2}$.} for a boson star with $\tilde{b}=0.1$ and the Hoffman's solution with the same $M$ and $Q$. 
 The asymptotic behaviour of the BI boson stars matches the electrovacuum solution in Einstein-BI theory.


\section{Conclusions}
\label{Sec:Conclusions}

The Born-Infeld theory of electrodynamics introduces a maximum strength for the electric and magnetic fields, which prevents the formation of singularities and makes the theory well-defined.
The Born-Infeld theory has been further developed and has found important applications in many areas of physics, including string theory, cosmology, and condensed matter physics. The importance of this theory is that it provides a more consistent and well-behaved framework for describing the behavior of electric and magnetic fields. 

As an application of this theory in a strong gravity scenario  
we have constructed in this work static spherically symmetric solutions of the Einstein field equations coupled to a complex scalar field and to a Born-Infeld electric field. These configurations, Born-Infeld boson stars, are quite similar to charged boson stars constructed in the past using Maxwell electrodynamics.

We found that Born-Infeld boson stars share some properties with their Maxwell counterparts, as the existence of a maximum mass given the central value of the scalar field, the existence of a limiting value of the charge of the field beyond which solutions can not be found and that the mass of the star increases as the charge increases. 

We also presented the space of solutions plotting the mass of the stars as a function of the frequency for representative values of $q$ and the Born-Infeld parameter $b$, we found that Born-Infeld boson stars are in general, less massive than pure Maxwell configurations.
We found that for a given value of $q$ the maximum possible mass of a boson star is greater in Maxwell than in Born-Infeld electrodynamics independently of the value of $b$.

We also showed that the spacetime of a Born-Infeld boson star asympotically matches the spacetime of the electrovacuum theory, that is Einstein field coupled to a pure electric Born-Infeld field.
This electrovacuum solution however, posses a event horizon and hence represent a charged black hole.

We showed that
the effect of the parameter $b$ in Born-Infeld boson stars is to decrease the intensity of the electric field throughout space. This explains the lower values for the total mass of the configurations and why it is possible to obtain solutions whose critical charge exceeds the Maxwell bound $q_\mathrm{crit}=1/\sqrt{2}$, for small values of $b$.

Finally, studying the stability of Born-Infeld boson stars can provide insights into the fundamental nature of the scalar field that composes the star and 
for gaining insights into fundamental physics. Such studies would be interesting to explore in a future work.


\acknowledgments
This work was partially supported by 
DGAPA-UNAM through grants IN110523 and IN105920, by the CONACyT Network Project No. 376127 ``Sombras, lentes y ondas gravitatorias generadas por objetos compactos astrof\'\i sicos" and No. 304001 ``Estudio de campos escalares con aplicaciones en cosmolog\'ia y astrof\'isica"
and by the European Union's Horizon 2020 research and innovation (RISE)
program H2020-MSCA-RISE-2017 Grant
No. FunFiCO-777740.
V. J. acknowledges financial support from CONACyT 
graduate grant program.  D. M. C. acknowledges financial support from CONACyT through a postdoctoral research grant.


\bibliography{ref}


\end{document}